\documentclass[fleqn,useAMS,usenatbib]{mnras}

\usepackage{amsmath}
\usepackage{amssymb}
\usepackage{graphicx}
\usepackage{times}
\usepackage{hyperref}
\usepackage{ulem}

\usepackage[flushleft]{threeparttable}

\title[Nuclear star formation and SMBH coalescence]
{RABBITS -- I. The crucial role of nuclear star formation in driving the coalescence of supermassive black hole binaries}

\author[S. Liao et al.]
{Shihong Liao,$^{1,2}$\thanks{Email: shliao@nao.cas.cn} Dimitrios Irodotou,$^{1}$ Peter H. Johansson,$^{1}$ Thorsten Naab,$^{3}$ \newauthor Francesco Paolo Rizzuto,$^{1}$ Jessica M. Hislop,$^{1}$ Alexander Rawlings,$^{1}$ Ruby J. Wright$^{1}$
\\
$^1$Department of Physics, University of Helsinki, Gustaf Hällströmin katu 2, FI-00014 Helsinki, Finland\\
$^2$Key Laboratory for Computational Astrophysics, National Astronomical Observatories, Chinese Academy of Sciences, Beijing 100101, China\\
$^3$Max-Planck-Institut f\"ur Astrophysik, Karl-Schwarzchild-Str 1, D-85748 Garching, Germany\\
}

\begin{document}



\maketitle

\label{firstpage}

\begin{abstract}
In this study of the `Resolving supermAssive Black hole Binaries In galacTic hydrodynamical Simulations' (RABBITS) series, we focus on the hardening and coalescing process of supermassive black hole (SMBH) binaries in galaxy mergers. For simulations including different galaxy formation processes (i.e. gas cooling, star formation, SMBH accretion, stellar and AGN feedback), we systematically control the effect of stochastic eccentricity by fixing it to similar values during the SMBH hardening phase. We find a strong correlation between the SMBH merger time-scales and the presence of nuclear star formation. Throughout the galaxy merging process, gas condenses at the centre due to cooling and tidal torques, leading to nuclear star formation. These recently formed stars, which inherit low angular momenta from the gas, contribute to the loss cone and assist in the SMBH hardening via three-body interactions. Compared to non-radiative hydrodynamical runs, the SMBH merger time-scales measured from the runs including cooling, stellar and SMBH physical processes tend to be shortened by a factor of ${\sim}$1.7. After fixing the eccentricity to the range of $e \sim 0.6$--$0.8$ during the hardening phase, the simulations with AGN feedback reveal merger time-scales of ${\sim} 100$--$500$ Myr for disc mergers and ${\sim} 1$--$2$ Gyr for elliptical mergers. With a semi-analytical approach, we find that the torque interaction between the binary and its circumbinary disc has minimal impact on the shrinking of the binary orbit in our retrograde galaxy merger. Our results are useful in improving the modelling of SMBH merger time-scales and gravitational wave event rates.
\end{abstract}

\section{Introduction}\label{sec:intro}

Supermassive black hole (SMBH, with mass in the range of $M_{\rm BH} \sim 10^5$--$10^{10} {\rm M}_{\sun}$) binaries within galaxy mergers are the primary target for low-frequency gravitational wave (GW) observatories. These include the ongoing nano-Hertz Pulsar Timing Array (see \citealt{Burke-Spolaor2019} for a review, and \citealt{Agazie2023Nanograv,Agazie2023,Antoniadis2023,Reardon2023,Xu2023} for recent results) and the upcoming milli-Hertz space-borne detectors such as the Laser Interferometer Space Antenna \citep[LISA,][]{Amaro-Seoane2017,Amaro-Seoane2023}, TianQin \citep{Luo2016}, and Taiji \citep{Ruan2020}. Understanding the coalescence process of SMBH binaries in detail is pivotal for predicting SMBH merger time-scales and event rates \citep{Begelman1980,Amaro-Seoane2023}.

During galaxy mergers, the central SMBHs initially sink to the remnant nucleus due to dynamical friction from stars, dark matter, and gas \citep{Chandrasekhar1943,Ostriker1999}, forming a gravitationally bound binary. The separation of the binary SMBHs continues to drop rapidly, initially because of dynamical friction and, later on, through three-body interactions with the surrounding stars, also referred to as gravitational slingshot interactions \citep[e.g.][]{Mikkola1992,Quinlan1996,Sesana2006}. When the semimajor axis of the SMBH binary reaches the hard binary separation (usually at ${\sim}$pc scales), where the specific binding energy of the binary equals the specific kinetic energy of stars, it forms a {\it hard} binary. The hard binary can eject nearby stars at high velocities beyond its influence radius (i.e. the radius within which the enclosed stellar mass is twice the binary mass), effectively excluding them from further interactions. The ejections of stars can lead to the formation of a large stellar density core in massive elliptical galaxies \citep[e.g.][]{Milosavljevic2001,Milosavljevic2002,Merritt2006,Rantala2018,Frigo2021,Nasim2021core}. Further shrinking of the binary orbit necessitates replenishing the loss cone, i.e. the low angular momentum region of the phase space where the stars are prone to interact with the binary \citep[see][for a review]{Merritt2013}. Whether the two SMBHs of a hard binary can reach small enough separations (i.e. $\la$ mpc) to enter the GW emission-driven phase \citep{Peters1963,Peters1964} has important implications for GW observations.

Previous gas-free $N$-body studies found that a hard SMBH binary might stall at a separation of roughly one parsec due to the depletion and inefficient refilling of the loss cone. This is referred to as the final parsec problem \citep{Milosavljevic2003}. This phenomenon was seen in idealized spherical systems, where the refilling process occurs through the two-body relaxation mechanism. This gives rise to a hardening time-scale that strongly depends on the number of star particles within the system, i.e. ${\sim} N$ \citep{Makino2004,Berczik2005}. If we extrapolate the result to the real galaxy systems in the Universe, where $N$ is large as it represents the number of actual stars, this time-scale can surpass the Hubble time, implying that the binary could fail to get through the final parsec and thus could not enter the GW-driven regime.

Several possible solutions have been proposed to overcome the final parsec problem bottleneck. Especially, departures from spherical symmetry have emerged as an effective solution to this problem \citep{Merritt2004,Berczik2006,Khan2011,Preto2011}. Specifically, in triaxial potentials, which are a natural outcome of galaxy mergers, there is a large population of stars with centrophilic orbits which can travel close to the centre and interact with the binary, facilitating the two SMBHs to enter the GW-driven phase. The Brownian motion, i.e. the random wandering of the binary's centre of mass (CoM) triggered by encounters with stars, might also help the binary to interact with more stars \citep{Quinlan1997,Merritt2001,Chatterjee2003,Bortolas2016}. The interaction with a third SMBH, introduced through a subsequent galaxy merger, can excite the binary eccentricity to high values \citep[e.g. the von Zeipel--Kozai--Lidov mechanism,][]{vonZeipel1910,Kozai1962,Lidov1962}, substantially reducing the merger time-scale and thus enhancing the SMBH merger rate \citep[e.g.][]{Blaes2002,Iwasawa2006,Hoffman2007,Kulkarni2012,Ryu2018,Bonetti2019,Mannerkoski2021}. Moreover, the presences of massive perturbers \citep[e.g.][]{Perets2008,Bortolas2018,ArcaSedda2019} and galaxy rotation \citep{Holley-Bockelmann2015,Mirza2017} can also contribute to the shrinking of a hard binary. It is important to note that real galaxy systems may experience a combination of these mechanisms concurrently.

Apart from the aforementioned dissipationless gravitational mechanisms, dissipative gas processes are also expected to affect the shrinking of the binary orbit, especially in gas-rich disc galaxy mergers. The gas dynamical friction and torques from circumnuclear discs have been suggested to play an important role in bringing two SMBHs to form a binary \citep[e.g.][]{Armitage2002,Escala2004,Escala2005,Dotti2006,Mayer2007,Chapon2013,SouzaLima2017}. During the hard binary phase, the SMBH binary further interacts with the gaseous circumbinary disc \citep[CBD, see][for a recent review]{Lai2022}. Small-scale hydrodynamical CBD simulations have revealed that the gas accretion from the CBD drives the binary toward equal mass \citep[e.g.][]{Artymowicz1996,Hayasaki2007,Farris2014,Duffell2020}. Furthermore, the torques from the CBD can lead to either a shrinkage or an expansion of the binary orbit, depending on the disc and binary properties \citep[e.g.][]{MacFadyen2008,Roedig2012,Miranda2017,Munoz2019,Munoz2020,Duffell2020,Heath2020,Tiede2020,Franchini2021,Franchini2023,Siwek2023,Tiede2023}. Note that the exact effects from CBDs remain a subject of ongoing debate.

In this work, we introduce the {\it `Resolving supermAssive Black hole Binaries In galacTic hydrodynamical Simulations'} (RABBITS) series of studies, aiming to investigate the orbital evolution of SMBH binaries down to the GW emission phase in galactic-scale hydrodynamical simulations of both disc and elliptical galaxy mergers. In a companion paper, \citet{Liao2023_PaperI} (hereafter Paper II), we demonstrate that the central properties of galaxies and the sinking and formation of SMBH binaries can be significantly influenced by galaxy formation processes, especially the active galactic nuclei (AGNs) feedback process and its specific numerical implementations.

In this study (Paper I of the RABBITS series), we focus on the evolution of SMBH binaries during the hardening and GW emission phases. We examine how various galaxy formation processes impact the binary hardening. Previous studies have shown that the eccentricity of an SMBH binary, when it becomes hard, exhibits stochastic behaviour due to the random encounters with stars \citep[e.g.][]{Quinlan1997,Berczik2005,Nasim2020,Gualandris2022,Rawlings2023}. This stochasticity in binary eccentricity can affect the estimation of the SMBH merger time-scale and may contaminate the effects of galaxy formation processes. In this study, we systematically mitigate the impact of stochastic eccentricities by fixing them to similar values in the different simulations.

The results in this work reveal a strong correlation between the merger time-scales and the nuclear star formation during the galaxy merger. This suggests another crucial process for accelerating the hardening of SMBH binaries, which is particularly important in gas-rich disc mergers.

This paper is organized as follows. In Section~\ref{sec:sim}, we describe the numerical code and the simulation details. The evolution of orbital parameters and the comparison of SMBH merger time-scales are presented in Section~\ref{sec:orb_params}. The physical processes driving SMBH coalescence are analysed and discussed in Section~\ref{sec:phys_proc}. We summarize and conclude in Section~\ref{sec:con}. Appendices \ref{ap:reset_ecc} and \ref{ap:res_study} offer details on the eccentricity resetting and the resolution study, respectively.

\section{Numerical simulations}\label{sec:sim}

As we focus on the evolution of SMBH binaries during the hardening and GW emission phases in this study, we initiate the simulations by selecting a snapshot from a parent galaxy merger simulation at the point where the SMBH binary becomes hard ($t = t_{\rm hard}$). Subsequently, the binary eccentricity is reset to a desired value, and the simulation is evolved forward until the two SMBHs eventually merge. In this section, we first provide an overview of our numerical code, then detail the information of the parent galaxy merger runs, and finally introduce the simulations that were rerun from $t_{\rm hard}$.

Following \citet{Merritt2013}, in this work, the hard binary separation is defined as 
\begin{equation}
    R_{\rm hard} = \frac{G M_{\rm red}}{4 \sigma_{\star}^2},
\end{equation}
where $G$ is the gravitational constant, $M_{\rm red} = M_{\rm BH,1} M_{\rm BH,2} / (M_{\rm BH,1} + M_{\rm BH,2})$ is the reduced mass of the SMBH binary, $M_{\rm BH, 1}$ and $M_{\rm BH,2}$ are the masses of the two SMBHs (assuming $M_{\rm BH, 2} \leq M_{\rm BH, 1}$), and $\sigma_{\star}$ is the one-dimensional stellar velocity dispersion within the half stellar mass radius. The hard binary separation is roughly the radius where the specific binding energy of the SMBH binary equals the specific kinetic energy of the surrounding stars. An SMBH binary is termed a hard binary when its semimajor axis becomes smaller than $R_{\rm hard}$.

\subsection{Numerical code}

In this work, the numerical code we used to perform the simulations is the \textsc{ketju} code \citep{Rantala2017}, which is based on the \textsc{gadget-3} code \citep[i.e. the successor of \textsc{gadget-2} described by][]{Springel2005} and utilizes the algorithmically regularized integrator \textsc{mstar} \citep{Rantala2020} to resolve the SMBH dynamics.\footnote{The public {\sc ketju} version, which is based on the {\sc gadget-4} code \citep{Springel2021}, can be found from \citet{Mannerkoski2023}. However, note that the public version does not contain the galaxy formation subgrid model used in this study.} To account for the orbital evolution of SMBH binaries due to GW emission, the {\sc ketju} code also includes the post-Newtonian (PN) corrections up to the order of 3.5PN for SMBH binaries \citep{Thorne1985,Mora2004,Blanchet2014}. To compute the gas dynamics, we use the \textsc{sphgal} smoothed particle hydrodynamics (SPH) implementation \citep{Hu2014} with the pressure-entropy formulation. We adopt the Wendland $C^4$ kernel \citep{Dehnen2012} and set the number of SPH neighbours to $N_{\rm ngb} = 100$. 

For the galaxy formation subgrid model including radiative cooling, star formation, and stellar feedback, the \textsc{ketju} code adopts the subgrid model which was originally developed by \citet{Scannapieco2005,Scannapieco2006} and later improved in \citet{Aumer2013} and \citet{Nunez2017}. This model has been used both in isolated and merger galaxy simulations \citep{Eisenreich2017,Liao2023} and in cosmological zoom-in simulations \citep{Mannerkoski2021,Mannerkoski2022}. The abundances of 11 chemical elements (i.e. H, He, C, N, O, Ne, Mg, Si, S, Ca, and Fe) are traced in the simulation, and the metal-dependent cooling rate tables are adopted from \citet{Wiersma2009}. The star formation conditions are as follows: the gas density $\rho_{\rm gas} \geq 2.2 \times 10^{-24}~{\rm g}~{\rm cm}^{-3}$ (i.e. the hydrogen number density $n_{\rm H} \geq 1~{\rm cm}^{-3}$), the gas temperature $T_{\rm gas} \leq 1.2 \times 10^{4}~{\rm K}$, and the gas flow should be convergent (i.e. the velocity divergence $\nabla \cdot \mathbfit{v}_{\rm gas} \leq 0$). When a gas particle satisfies these criteria, it is stochastically converted into a star particle. For stellar feedback, we consider the effects from both Type Ia and Type II supernova explosions and the winds of asymptotic giant branch (AGB) stars. When stellar feedback is performed, star particles give masses (i.e. metals) and thermal and kinetic energy to their surrounding gas particles. We refer the reader to \citet{Nunez2017} for more details of the stellar feedback model.

For SMBH accretion, we use the SMBH binary accretion subgrid model introduced in \citet{Liao2023}, which extends the widely used Bondi--Hoyle--Lyttleton (BHL) accretion \citep{Hoyle1939,Bondi1944,Bondi1952} model into the SMBH binary phase and incorporates the preferential accretion subgrid model of CBDs \citep[see][for a recent review]{Lai2022}. Specifically, when two SMBHs form a gravitationally bound binary, the total accretion rate is determined using the BHL formula with the gas properties computed at the binary's CoM position. Then, the gas accretion on to each SMBH is distributed based on the fitting formula provided by \citet{Duffell2020}. As a result, the secondary SMBH (i.e. the one with the lower mass) experiences a higher accretion rate, facilitating the evolution of the binary towards equal masses. This binary accretion model provides more physically motivated SMBH mass evolution in the binary phase, which is important in modelling the AGN feedback and the GW induced recoil velocities. 

To study the impact of different AGN feedback implementations, we consider both pure thermal feedback and pure kinetic feedback. For the thermal AGN feedback model, we follow the implementation of \citet{Springel2005feedback}. The thermal feedback is continuously performed at each time-step when the SMBH particle is active. Specifically, the amount of feedback energy is first computed according to the SMBH accretion rate, then this energy is added to the internal energy of the surrounding gas particles according to their SPH kernel weights. For the kinetic AGN feedback model, we adopt an implementation similar to the kinetic feedback mode proposed in \citet{Weinberger2017}. In contrast to the aforementioned thermal AGN feedback, the kinetic AGN feedback here is implemented using a pulsed approach, i.e. at each time-step when the SMBH is active, the feedback energy -- derived from the SMBH accretion rate -- is added to a feedback energy reservoir, and the energy in the reservoir is released only after it reaches a predefined threshold. When the release occurs, the surrounding gas particles get kicked and their kinetic energies are boosted. The direction of the kick velocity is equally probable (50 per cent chance) to be either parallel or anti-parallel to either the angular momentum direction of the kicked gas particle with respect to the SMBH (in the single SMBH phase) or the direction of the orbital angular momentum of the SMBH binary (in the binary SMBH phase). More details on the code implementation and the adopted parameter values can be found in \citet{Liao2023}.

\subsection{Parent runs}

\begin{table*}
\begin{threeparttable}
\caption{Simulation information. From left to right, the galaxy merger, the simulation set, the included physical processes, the time when the SMBH binary becomes hard ($t_{\rm hard}$), the eccentricity at $t_{\rm hard}$ from the parent simulation, the number of star particles at $t_{\rm hard}$, the number of gas particles at $t_{\rm hard}$, and the number of rerun realizations. The abbreviations for the physical processes: Grav -- gravity, SPH -- smoothed particle hydrodynamics, Cool -- gas cooling, Star -- star formation and stellar feedback, ThmAGN -- SMBH accretion and thermal AGN feedback, KinAGN -- SMBH accretion and kinetic AGN feedback.}
\label{tab:sim_info}
\begin{tabular}{lllccccccc}
\hline
Galaxy & Simulation set & Included physics & $t_{\rm hard}$ & $e_{\rm hard}^{\rm parent}$ & $N_{\star}$ & $N_{\rm gas}$ & Number of rerun & \\
merger & & & [Gyr] &  & $[10^{6}]$ & $[10^{6}]$ & realizations & \\
\hline
DD-11-G5 & NoGas & Grav & 1.76 & 0.492 & 2.87 & 0 & 10 & \\
DD-11-G5 & NoCool & Grav, SPH & 1.71 & 0.995 & 2.44 & 0.43 & 10 & \\
DD-11-G5 & CoolStarNoAGN & Grav, SPH, Cool, Star & 1.56 & 0.931 & 2.77 & 0.11 & 10 & \\
DD-11-G5 & CoolStarThmAGN & Grav, SPH, Cool, Star, ThmAGN & 1.58 & 0.624 & 2.78 & 0.11 & 10 & \\
DD-11-G5 & CoolStarKinAGN & Grav, SPH, Cool, Star, KinAGN & 1.64 & 0.638 & 2.68 & 0.19 & 10 & \\
\\
EE-11-G5 & NoGas & Grav & 1.95 & 0.885 & 2.96 & 0 & 10 & \\
EE-11-G5 & NoCool & Grav, SPH & 2.00 & 0.935 & 2.44 & 0.53 & 10 & \\
EE-11-G5 & CoolStarNoAGN & Grav, SPH, Cool, Star & 1.80 & 0.792 & 2.46 & 0.51 & 10 & \\
EE-11-G5 & CoolStarThmAGN & Grav, SPH, Cool, Star, ThmAGN & 1.90 & 0.372 & 2.45 & 0.51 & 10 & \\
EE-11-G5 & CoolStarKinAGN & Grav, SPH, Cool, Star, KinAGN & 2.00 & 0.921 & 2.44 & 0.53 & 10 & \\
\hline
\end{tabular}
\end{threeparttable}
\end{table*}

The parent galaxy merger simulations are described and analysed in detail in Paper II, consisting of two equal-mass galaxy mergers: the disc-disc merger `DD-11-G5' and the elliptical-elliptical merger `EE-11-G5'. These simulation names follow the convention of `progenitor galaxy types-galaxy mass ratio-orbit geometry' introduced in \citet{Liao2023}. The so-called G5 retrograde orbit, adopted from \citet{Naab2003}, is specified by the inclination angles between the spin plane and the orbit plane ($i$) and the arguments of pericentre ($\omega$) for the two galaxies \citep{Toomre1972}: $i_{1} = -109^\circ$, $\omega_{1} = -60^\circ$, $i_{2} = 180^\circ$, $\omega_{2} = 0^\circ$. The G5 orbit geometry is adopted as it exhibits modest starbursts during the merger, which helps to make the simulations computationally efficient.

The initial disc and elliptical galaxies have similar masses in their different components (stars, gas, and dark matter) but differ in their morphology. Specifically, the initial disc galaxy consists of a stellar disc, a stellar bulge, a gas disc, a central SMBH, and a dark matter halo. Conversely, the initial elliptical galaxy is composed of a stellar bulge, a hot gas halo, a central SMBH, and a dark matter halo. For both galaxies, the masses for the dark matter, stellar, and gaseous components, and the SMBH are $M_{\rm DM} = 2.5 \times 10^{12}~{\rm M}_{\sun}$, $M_{\star} = 1.2 \times 10^{11}~{\rm M}_{\sun}$, $M_{\rm gas} \approx 2.5 \times 10^{10}~{\rm M}_{\sun}$, and $M_{\rm BH} = 7.5 \times 10^{7}~{\rm M}_{\sun}$, respectively.

The initial particle masses of dark matter, stars, and gas are $m_{\rm DM} = 1.6 \times 10^{6}~{\rm M}_{\sun}$, $m_{\star} = 10^{5}~{\rm M}_{\sun}$, and $m_{\rm gas} = 10^{5}~{\rm M}_{\sun}$, respectively. The numbers of particles for different galaxy components in the initial conditions are $N_{\rm DM} = 3.2 \times 10^{6}$, $N_{\star} = 2.4 \times 10^{6}$, and $N_{\rm gas} \approx 5 \times 10^{5}$. The softening lengths adopted in the simulations are $\epsilon_{\rm DM} = 100$ pc, $\epsilon_{\star} = \epsilon_{\rm BH} = 5$ pc, and $\epsilon_{\rm gas} = 20$ pc.

For both the DD-11-G5 and EE-11-G5 galaxy mergers, we perform simulation sets including different physical processes:
\begin{enumerate}
\item {\bf NoGas}. This is a pure gravity simulation, with all gas particles in the initial condition being converted into star particles.
\item {\bf NoCool}. Compared to the NoGas case, this run includes SPH computations for gas particles with an adiabatic equation of state.
\item {\bf CoolStarNoAGN}. Compared to NoCool, this run further includes gas cooling, star formation, and stellar feedback. However, no SMBH accretion and AGN feedback processes are considered in this run.
\item {\bf CoolStarThmAGN}. Compared to CoolStarNoAGN, this run further includes SMBH accretion and the pure thermal AGN feedback process. 
\item {\bf CoolStarKinAGN}. In contrast to CoolStarThmAGN, this run incorporates pure kinetic AGN feedback instead.
\end{enumerate}
In total, the parent runs comprises ten simulations. Each simulation starts from the cosmic time of $t_0 = 10.7$ Gyr, and is evolved for a duration of 3 or 4 Gyr (depending on when the two SMBHs finally merge). As we show in Paper II, the properties of our initial galaxies and the galaxy merger remnants from our simulations which include the AGN feedback process (i.e. CoolStarThmAGN and CoolStarKinAGN) are in good agreement with the observations.

For different parent runs, we summarize in Table~\ref{tab:sim_info} the simulation time\footnote{The simulation time is defined as the time span after $t_{0}$.} ($t_{\rm hard}$), the binary eccentricity ($e_{\rm hard}^{\rm parent}$), and the number of star ($N_{\star}$) and gas particles ($N_{\rm gas}$) from the first snapshot at which the two SMBHs reach the hard binary phase. We refer the interested reader to Paper II for more details and analyses of the parent runs.

\subsection{Reruns with $e_{\rm hard} = 0.6$}

The eccentricity of a binary is a key parameter that strongly influences the SMBH merger time-scales. According to \citet{Peters1964}, the rates of decay for the semimajor axis ($a$) and the eccentricity ($e$) due to GW emission can be approximated by (i.e. at the 2.5PN level)
\begin{equation}
    \frac{{\rm d} a}{{\rm d} t} = - \frac{64}{5} \frac{G^3 M_{\rm BH,1} M_{\rm BH,2} M_{\rm bin}}{c^5 a^3 \left(1 - e^2\right)^{7/2}} \left(1 + \frac{73}{24}e^2 + \frac{37}{96}e^4\right),
\end{equation}
\begin{equation}
    \frac{{\rm d} e}{{\rm d} t} = - \frac{304}{15} \frac{G^3 M_{\rm BH,1} M_{\rm BH,2} M_{\rm bin}}{c^5 a^4 \left(1 - e^2\right)^{5/2}} e \left(1 + \frac{121}{304}e^2\right).
\end{equation}
Here, $M_{\rm bin} = M_{\rm BH,1} + M_{\rm BH,2}$ is the total SMBH mass, $c$ is the speed of light. When the eccentricity of a binary is closer to 1, the binary enters the GW emission-dominated phase earlier due to the factor of $(1-e^2)^{-7/2}$, resulting in a shorter SMBH merger time-scale. In the parent simulations, distinct runs exhibit quite different eccentricities at the time when the two SMBHs become hard (see the column of $e_{\rm hard}^{\rm parent}$ in Table~\ref{tab:sim_info}). Previous numerical simulations have demonstrated that the eccentricity of a binary at $t_{\rm hard}$ displays inherent stochasticity, arising from random stellar encounters \citep[e.g.][]{Quinlan1997,Berczik2005,Nasim2020,Gualandris2022,Rawlings2023}. This imposes a limitation on directly comparing SMBH merger time-scales across different parent runs, as well as the study of how factors beyond GW emission influence these time-scales.

To mitigate the impact from the stochastic nature of eccentricity, we manually reset $e_{\rm hard}$ to the same value in different runs, and rerun the simulations starting from $t_{\rm hard}$ and continuing until the point of final SMBH merger. To reset $e_{\rm hard}$, we take the snapshot of $t = t_{\rm hard}$ from each simulation, and keep the SMBH mass, the binary CoM position and velocity, and the semimajor axis unchanged. The coordinates and velocities of the two SMBHs are adjusted to match the specific relative angular momentum corresponding to the given new eccentricity. The eccentricity of a binary is determined by the magnitude of the specific relative angular momentum, therefore, we can randomly assign different directions to the specific relative angular momentum (or equivalently the orbital plane of the binary) to generate different realizations. The details of how the eccentricity is reset can be found in Appendix~\ref{ap:reset_ecc}.

\begin{figure*} 
\centering\includegraphics[width=\textwidth]{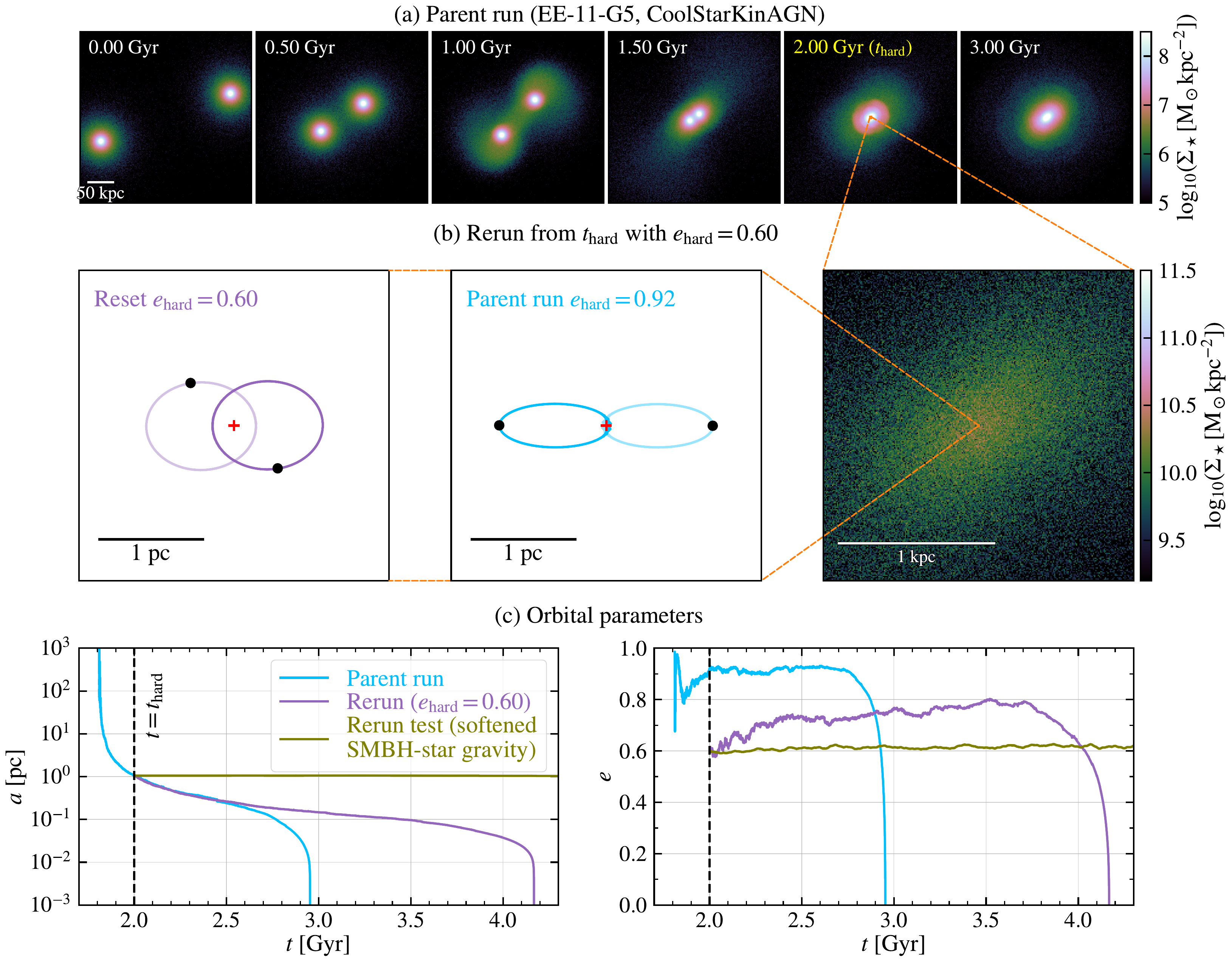}
\caption{Overview of the simulations used in this study. (a) Evolution of the stellar components in the CoolStarKinAGN parent run of the EE-11-G5 galaxy merger. (b) The right panel displays the projected stellar density in the $2 \times 2$ kpc region centered on the SMBH binary at $t = t_{\rm hard}$. In the middle panel, the two SMBHs are represented by black filled circles, and their first few orbits after the time $t_{\rm hard}$ from the parent run are shown. The red cross indicates the centre of mass of the binary. In this panel, the SMBH orbits are depicted on the orbital plane. The left panel illustrates the SMBHs and their first few orbits from the rerun, where the binary eccentricity was reset to $0.6$. (c) Evolution of the PN-corrected orbital parameters in the parent run (light blue), the rerun (purple), and a test rerun with softened gravity between the SMBH and star particles (olive). The comparison between the rerun and the test rerun clearly shows that in the absence of strong interactions with stars (i.e. softened gravity), the binary stalls at ${\sim}1$ pc and its eccentricity remains constant.}
\label{fig:sim_overview}
\end{figure*}

\begin{figure*} 
\centering\includegraphics[width=450pt]{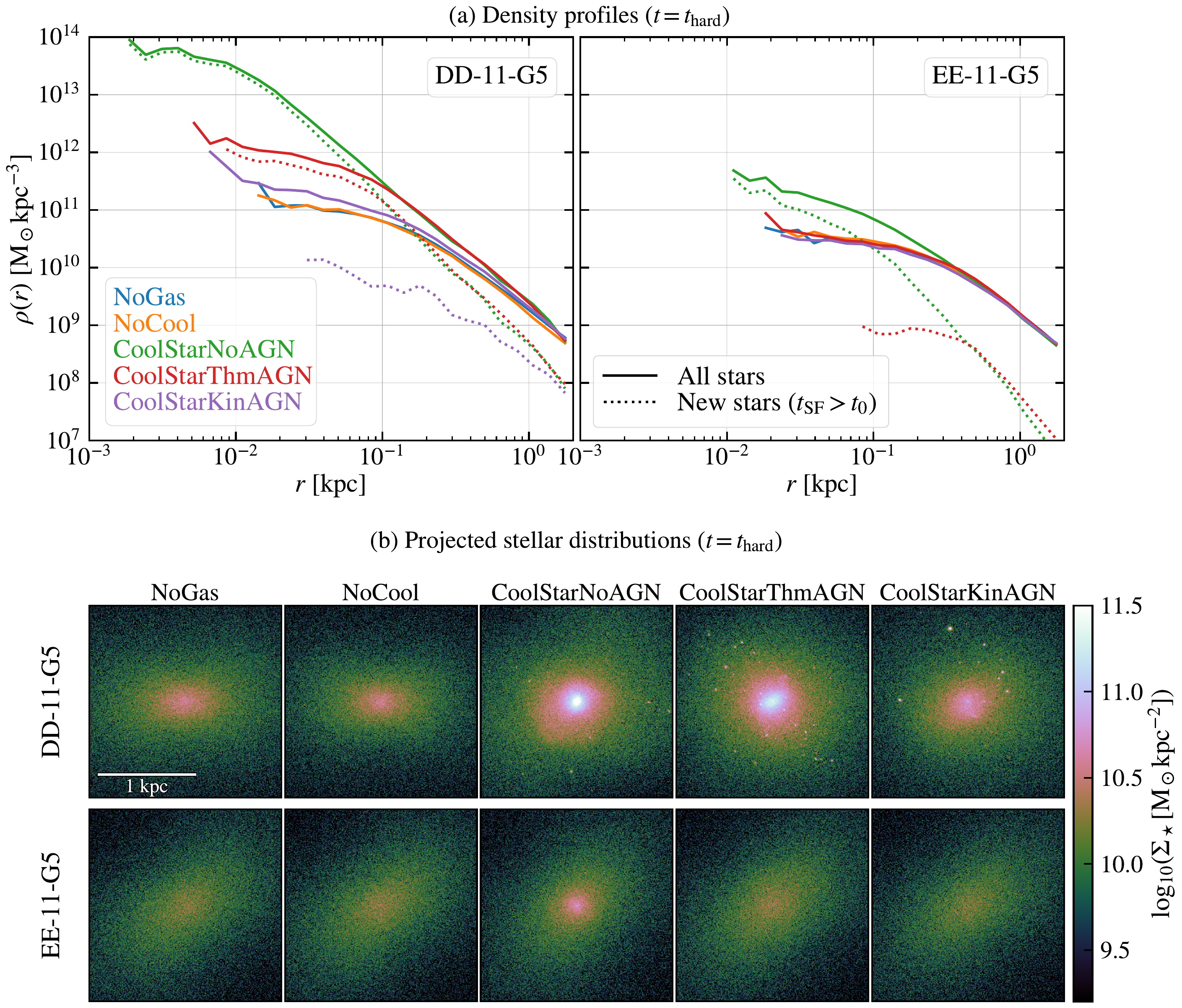}
\caption{Stellar distributions of galaxy remnants at $t = t_{\rm hard}$ from simulations including different galaxy formation processes. (a) Spherically averaged stellar density profiles at $t = t_{\rm hard}$ from the DD-11-G5 (left) and EE-11-G5 (right) runs. These profiles are centred on the CoM of the binary. The NoGas, NoCool, CoolStarNoAGN, CoolStarThmAGN, and CoolStarKinAGN runs are displayed with blue, orange, green, red, and purple lines, respectively. The solid lines plot the total stellar density profiles, while the dotted lines show the density profiles of newly formed stars. Here, the new stars are defined as the stars formed after the start of the parent simulations, i.e. with star formation time $t_{\rm SF} > t_{0}$. (b) The $2 \times 2$ kpc projected stellar distributions (on the $xy$-plane) centred on the SMBH binary from the DD-11-G5 (upper) and EE-11-G5 (lower) runs. From left to right, the NoGas, NoCool, CoolStarNoAGN, CoolStarThmAGN, and CoolStarKinAGN runs are plotted.}
\label{fig:thard_rho_star}
\end{figure*}

In this study, we opt to reset $e_{\rm hard}$ to an intermediate value of 0.6. This choice ensures that the GW emission dominated phase is neither triggered prematurely (making it challenging to disentangle the effects of eccentricity and GW emission from those ascribable to galaxy formation processes) nor delayed excessively (resulting in a longer time for the binary to merge and becoming computationally inefficient), making it well-suited for this pilot investigation. 

Once the initial condition for a rerun is set up, we use exactly the same numerical code and parameters as in the original parent run to evolve the simulation until the SMBHs merge. For each parent run, we conduct ten rerun realizations (i.e. they differ in the directions of the binary's specific relative angular momentum) to further minimize the influence of the eccentricity stochasticity on our conclusions. This approach enables us to estimate both the mean and standard deviation of the SMBH merger time-scales, facilitating direct comparisons across different simulations.

\section{Orbital parameters and merger time-scales} \label{sec:orb_params}

\subsection{Simulation overview}\label{subsec:overview}

In Fig.~\ref{fig:sim_overview}, we use the EE-11-G5 CoolStarKinAGN run as an illustrative example to provide an overview of our simulations. In row (a), we plot the evolution of the stellar components from the parent run. At around $t_{\rm hard} \approx 2.0$ Gyr, the SMBH binary in the centre of the galaxy remnant reaches the hard binary separation, and its eccentricity is then $e_{\rm hard}^{\rm parent} = 0.92$; see the middle and right panels in row (b). For the rerun, we take the snapshot at $t = t_{\rm hard}$ from the parent run, and then manually reset the binary eccentricity to $e_{\rm hard} = 0.6$ (the left panel of row b), after which we evolve the simulation until the point of the SMBH merger.

In row (c) of Fig.~\ref{fig:sim_overview}, we compare the evolution of the orbital parameters from the parent run (light blue) and one of the reruns (purple). As we reduce $e_{\rm hard}$ from 0.92 to 0.60, the SMBH merger time-scales increase from ${\sim} 1$ to ${\sim} 2$ Gyr. We notice that although we reset the initial eccentricity to a value of 0.6 in the rerun, the eccentricity gradually evolves to a higher value of ${\sim} 0.8$, before the GW emission starts to dominate and circularizes the binary. This originates from the three-body interactions between the binary and star particles. Similar phenomenon has been observed in previous pure $N$-body simulations \citep[e.g.][]{Berczik2005,Merritt2007}, in which a circular binary is initially placed at the centre of an isolated galaxy and the interactions with the passing stars rapidly induce nonzero eccentricities. It has also been shown in the three-body scattering experiments that the eccentricities in the stellar hardening phase tend to grow slightly over time \citep[e.g.][]{Mikkola1992,Quinlan1996,Sesana2006}. The rotation of the galaxy remnant can further shape the eccentricity evolution, i.e. the counter-rotation (co-rotation) between the stellar nucleus and the SMBH binary can lead to an increase (decrease) of the binary eccentricity \citep{Sesana2011}. However, we anticipate a minimal role of rotation in this run, given that the dimensionless spin parameter \citep{Bullock2001} of the stellar nucleus at $t_{\rm hard}$ is relatively low, e.g. within 1 kpc (500 pc), it is $\lambda_\star = 0.0016~(0.0022)$.\footnote{Similar conclusions apply to other runs. For the other simulations of the EE-11-G5 merger, the mean spin parameter is $\bar{\lambda}_\star = 0.005$, while for the DD-11-G5 mergers, it is $\bar{\lambda}_\star = 0.029$.} For comparison, the typical spin parameters of random motion-supported dark matter haloes and rotation-supported galactic discs are ${\sim}0.05$ and ${\sim}0.4$, respectively \citep{Mo2010}.

To further validate this stellar-driven explanation, we conducted a test using the same initial condition as the rerun, but this time incorporating softened gravity between the SMBH and star particles. Note that the interactions between SMBHs remained the same as in the rerun, i.e. non-softened gravity with PN corrections up to the order of 3.5PN. The evolution of orbital parameters in this test run is displayed in row (c) using an olive colour. It is evident that in the absence of strong interactions with stars (i.e. softened gravity), the binary stalls at ${\sim} 1$ pc. In addition, the binary eccentricity remains a constant around 0.6, providing further support for the stellar-driven eccentricity evolution.

\subsection{Star formation and central stellar density at $t_{\rm hard}$}

In Fig.~\ref{fig:thard_rho_star}, we show the stellar distributions of galaxy merger remnants at $t_{\rm hard}$, which are the initial conditions for the reruns. The spherically averaged stellar density profiles centred on the binary CoM are displayed in row (a), with solid and dotted lines showing the results from all stars and new stars, respectively. Here, the new stars are defined as those formed after the onset of the parent simulations, i.e. stars with a formation time $t_{\rm SF} > t_{0}$.

It is obvious that the central stellar densities are significantly affected by the formation of new stars, i.e. with more nuclear star formation, the central stellar density is markedly boosted. This is particularly evident in the disc-disc galaxy merger runs, where the absence of AGN feedback in the CoolStarNoAGN run leads to unrealistically high star formation in the centre. Once thermal AGN feedback is included, the central stellar density ($r \la 0.1$ kpc) decreases, but it is still about one order of magnitude higher than in the NoGas and NoCool cases which do not include star formation. The kinetic AGN feedback is more effective in suppressing star formation than the thermal case, resulting in a moderate central stellar density in the CoolStarKinAGN run.

In the elliptical-elliptical galaxy merger, the CoolStarNoAGN run also exhibits a large fraction of nuclear star formation, as there is no AGN feedback to prevent the cooling flows. Once AGN feedback is included, the central densities of the new stars are significantly suppressed. Especially for the case of kinetic AGN feedback, as shown in Paper II, the star formation is completely shut down roughly after the first pericentre passage of the two galaxies. Consequently, all runs except CoolStarNoAGN exhibit very similar stellar density profiles.

The differences in the central stellar distributions are also evident from the projected plots in row (b). Moreover, as shown in Paper II, the central mass distributions at $t_{\rm hard}$ in all runs exhibit clear triaxiality, a natural outcome of galaxy mergers. Compared to spherical potentials, triaxial potentials can assist the binary hardening process by providing more stars with centrophilic orbits, thus avoiding the final parsec problem \citep{Merritt2004,Berczik2006,Khan2011,Preto2011}.

From the comparisons above, we can conclude that nuclear star formation plays an important role in setting the environment in which the hard SMBH binary evolves.

\subsection{SMBH merger time-scales from reruns}

\begin{figure*} 
\centering\includegraphics[width=400pt]{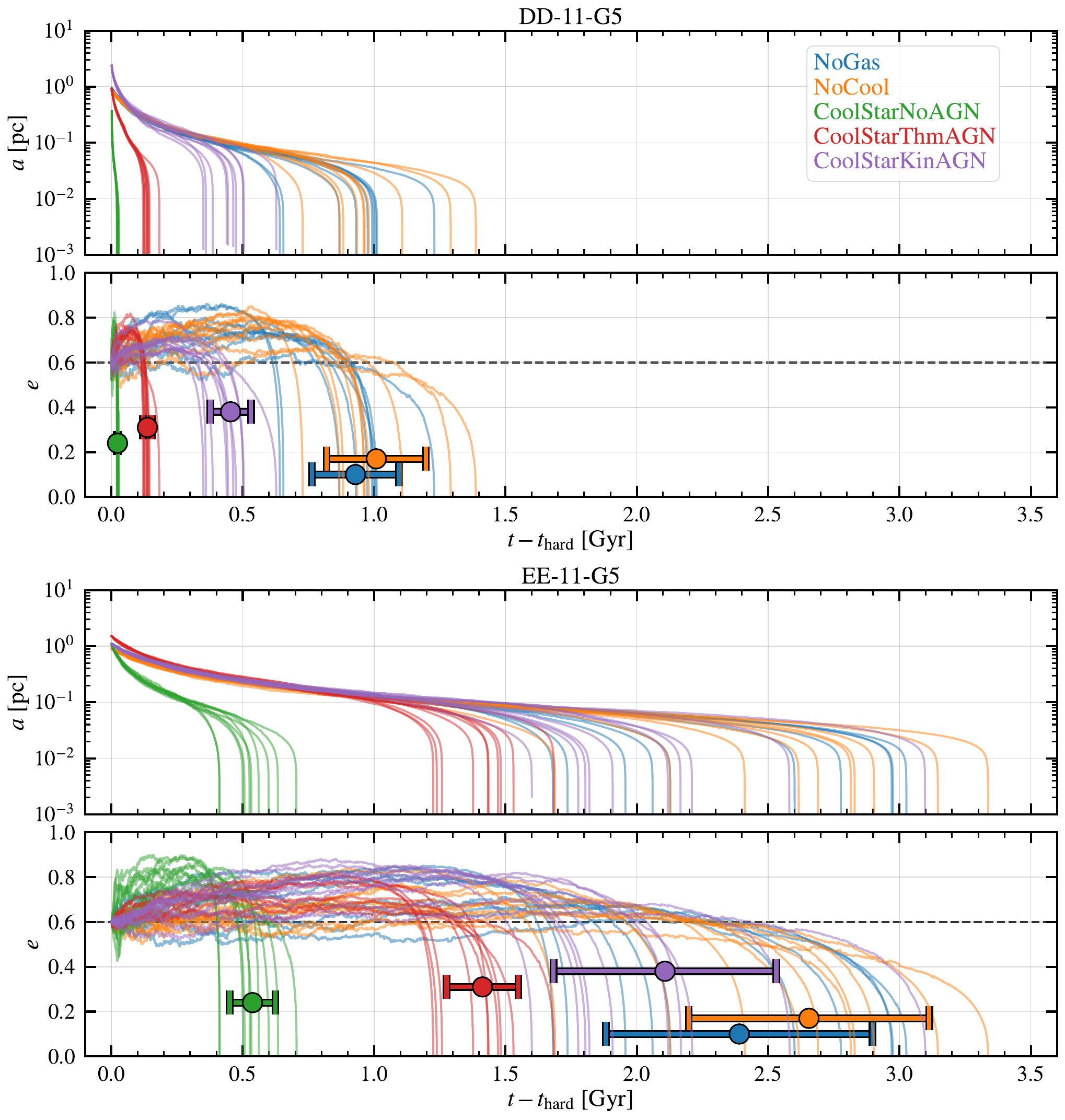}
\caption{Time-evolution of the orbital parameters in simulations that start from snapshots when the two SMBHs become dynamically hard (with the initial orbital eccentricity reset to 0.6). The NoGas, NoCool, CoolStarNoAGN, CoolStarThmAGN, and CoolStarKinAGN runs are shown with blue, orange, green, red, and purple colours, respectively. For each simulation set, all ten realizations are shown here. In the eccentricity panels, the filled circles and error bars mark the mean and the standard deviation of the SMBH coalescence time from ten realizations, and they are vertically shifted for different simulation sets in order to provide a clear illustration. The horizontal dashed lines mark $e = 0.6$.}
\label{fig:orb_params_fix_ecc}
\end{figure*}

\begin{figure} 
\centering\includegraphics[width=\columnwidth]{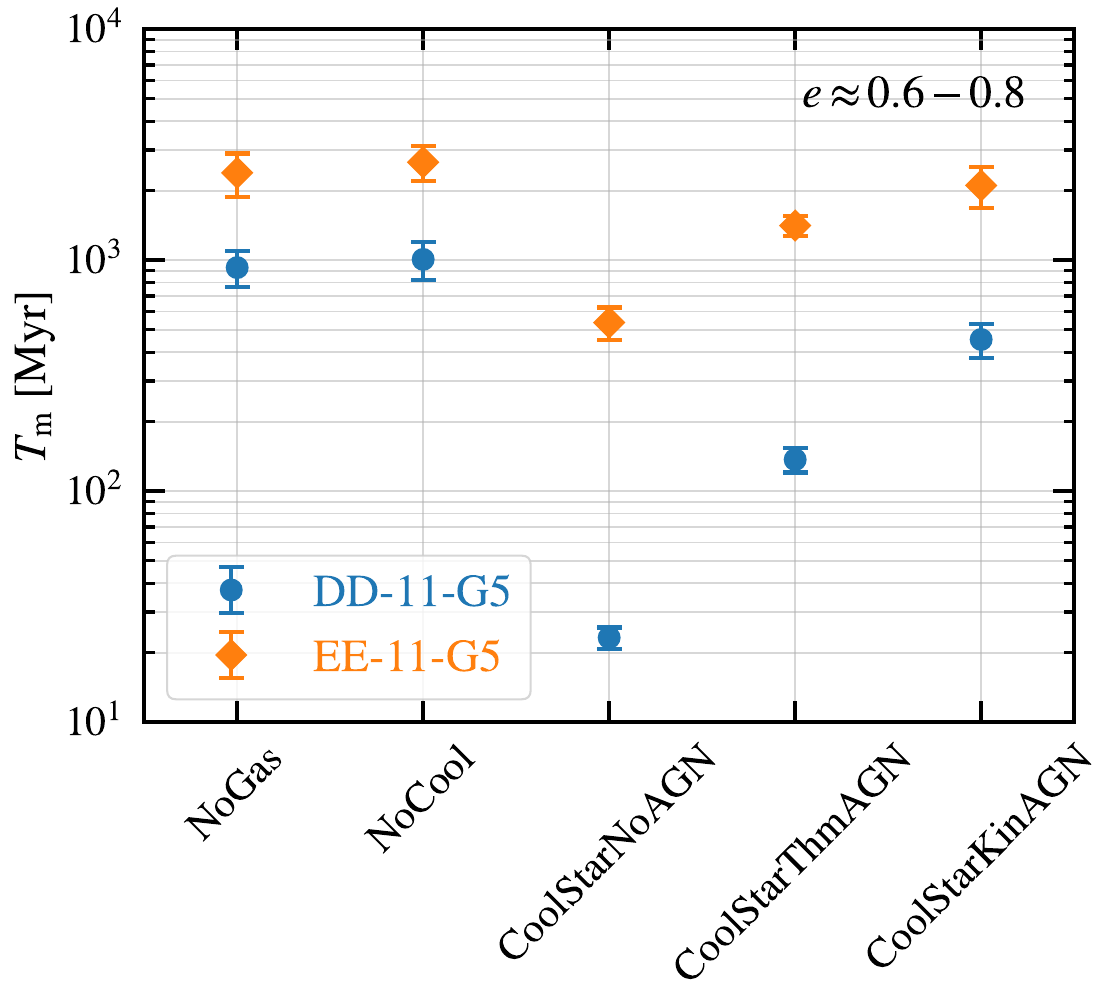}
\caption{SMBH merger time-scales from different runs. The blue circles and the orange diamonds show the mean merger time-scales (averaged over ten realizations) from DD-11-G5 and EE-11-G5 runs, respectively. For each set of runs, the error bar shows the standard deviation of the merger time-scales.}
\label{fig:merger_timescales}
\end{figure}

\begin{table}
\begin{threeparttable}
\caption{Merger time-scales (mean $\pm$ standard deviation) from the reruns. The values are computed from ten realizations.}
\label{tab:merger_timescale}
\begin{tabular}{lcc}
\hline
Simulation set & DD-11-G5 & EE-11-G5 \\
 & $T_{\rm m}$ [Myr] & $T_{\rm m}$ [Myr]\\
\hline
NoGas & 929.2 $\pm$ 165.7 & 2390.0 $\pm$ 507.4 \\
NoCool & 1008.3 $\pm$ 189.3 & 2655.3 $\pm$ 457.9 \\
CoolStarNoAGN & 23.2 $\pm$ 2.5 & 537.3 $\pm$ 87.0 \\
CoolStarThmAGN & 137.0 $\pm$ 16.5 & 1412.5 $\pm$ 137.0 \\
CoolStarKinAGN & 453.8 $\pm$ 77.5 & 2107.2 $\pm$ 424.1 \\
\hline
\end{tabular}
\end{threeparttable}
\end{table}

The time-evolution of the orbital parameters from all the reruns is summarized in Fig.~\ref{fig:orb_params_fix_ecc}. Each simulation set is represented by a distinct colour, and within each set, the results of all ten realizations are plotted.

Although all reruns start with an eccentricity of $e_{\rm hard} = 0.6$, the interactions with stellar particles result in varying eccentricities among the different realizations, as discussed in Section~\ref{subsec:overview}. Prior to the GW emission dominated phase, which ultimately circularizes the binary, the eccentricity spread in each simulation set falls approximately within the range of $e \sim 0.6$--$0.8$. This consistency in the eccentricity range allows us to alleviate the impact of eccentricity stochasticity and directly compare the mean SMBH merger time-scales across different simulation sets. In Fig.~\ref{fig:orb_params_fix_ecc}, the mean and standard deviation of the SMBH coalescence time, computed from ten realizations, are indicated by filled circles and error bars for different simulation sets. The detailed SMBH merger time-scales, $T_{\rm m} = t_{\rm coal} - t_{\rm hard}$ (with $t_{\rm coal}$ being the SMBH coalescence time), are summarized and compared in Table~\ref{tab:merger_timescale} and Fig.~\ref{fig:merger_timescales}.

In the case of the disc-disc galaxy merger, the NoGas and NoCool runs yield similar SMBH merger time-scales of ${\sim} 1$ Gyr. However, when additional galaxy formation processes are included, the merger time-scales exhibit a clear dependence on the central stellar density, which is affected by the AGN feedback process. In the absence of AGN feedback, the central stellar density is markedly boosted, and the SMBH merger time-scale is notably short, i.e. only ${\sim} 20$ Myr. When thermal AGN feedback is introduced, the central stellar density is lowered, and this time-scale is extended to ${\sim} 140$ Myr. With the more effective kinetic AGN feedback, the merger time-scale further increases to ${\sim} 450$ Myr.

For the elliptical-elliptical galaxy merger, the CoolStarNoAGN run reveals the shortest merger time-scale of ${\sim} 540$ Myr, followed by the CoolStarThmAGN run with $T_{\rm m} \sim 1.4$ Gyr. The other runs produce merger time-scales that are similar within the $1\sigma$ standard deviation, i.e. ${\sim} 2$ Gyr. Kinetic AGN feedback proves more effective in maintaining red and quiescent elliptical galaxies (as shown in Paper II). The similarity in SMBH merger time-scales between the CoolStarKinAGN and NoGas runs supports the gas-free assumption, which has been adopted in many previous studies of elliptical galaxies \citep[e.g.][]{Quinlan1997,Milosavljevic2001,Khan2011,Rantala2019,Frigo2021,Nasim2021core}.

Across all simulation sets, the SMBH merger time-scales display a consistent pattern, with gas-rich disc galaxy mergers exhibiting shorter time-scales than their gas-poor elliptical counterparts. Specifically, in the simulations with AGN feedback, $T_{\rm m}$ is ${\sim} 100$--$500$ Myr in disc galaxy mergers, while it is ${\sim } 1$--$2$ Gyr for elliptical galaxy mergers. It is important to keep in mind that these values are based on the reruns with eccentricities in the range of ${\sim} 0.6$--$0.8$ (prior to the GW emission-dominated phase). The quantitative SMBH merger time-scales will vary if we reset $e_{\rm hard}$ to other values in the rerun experiments.

By comparing the SMBH merger time-scales in Fig.~\ref{fig:merger_timescales} to the central stellar densities illustrated in Fig.~\ref{fig:thard_rho_star}, we notice a pronounced correlation between the SMBH merger time-scales and the nuclear stellar densities, i.e. higher nuclear stellar densities correspond to shorter SMBH merger time-scales. This is present in all simulation sets (including both DD-11-G5 and EE-11-G5 runs). As the variation of the nuclear stellar density here is primarily driven by star formation, this correlation hints that nuclear star formation plays an important role in influencing the binary hardening, especially in gas-rich disc galaxy mergers. We will scrutinize this picture quantitatively in the following section.

Finally, we note that we have performed a resolution study based on the CoolStarKinAGN run. This involved both increasing the mass resolution by a factor of two and decreasing the mass resolutions by factors of two and four (see Appendix~\ref{ap:res_study} for details). We found that for the DD-11-G5 runs, there exists a mild resolution dependence for the SMBH merger time-scales, where higher mass resolutions correspond to slightly shorter $T_{\rm m}$. However, the merger time-scale from the runs presented above agrees well with that from the higher resolution runs. For the EE-11-G5 runs, no resolution dependence is observed for the merger time-scales.

\section{Physical processes driving SMBH hardening} \label{sec:phys_proc}

\begin{figure*} 
\centering\includegraphics[width=500pt]{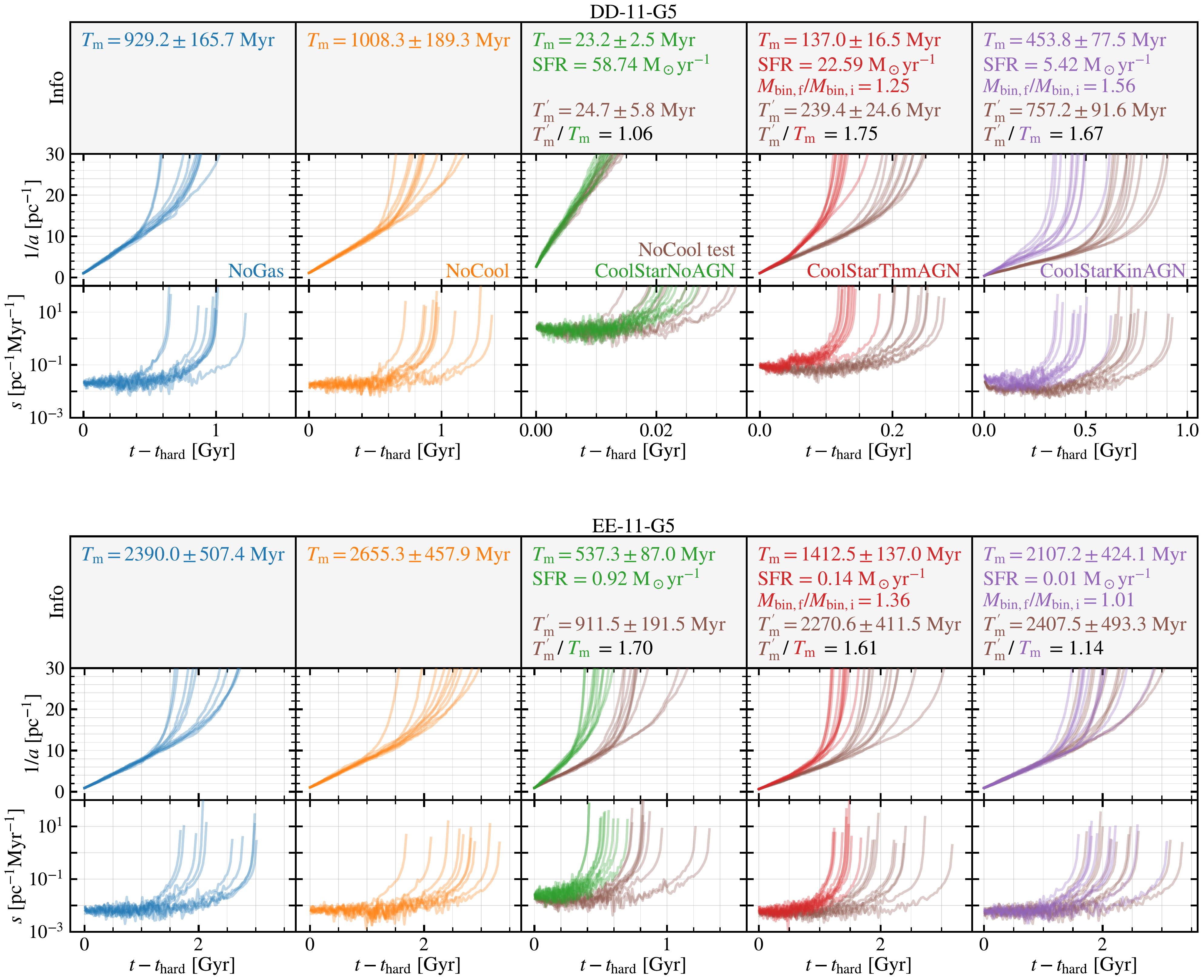}
\caption{Time evolution of the inverse semimajor axis ($1/a$) and hardening rate ($s$) from the DD-11-G5 (upper) and EE-11-G5 runs (lower). From left to right, all ten realizations of the NoGas (blue), NoCool (orange), CoolStarNoAGN (green), CoolStarThmAGN (red), and CoolStarKinAGN (purple) simulation sets are plotted. For each simulation set, the top panel summarizes the following information: the mean and standard deviation of the merger time-scales $T_{\rm m}$, the averaged SFR over $T_{\rm m}$ (if star formation is included), and the averaged SMBH mass growth factor $M_{\rm bin, f}/M_{\rm bin, i}$ (if SMBH accretion is included); the middle and bottom panels plot $1/a$ and $s$, respectively. In the CoolStarNoAGN, CoolStarThmAGN, and CoolStarKinAGN panels, the brown lines show the results from the NoCool tests, which start from the corresponding initial conditions but evolve with gravity and SPH only. The merger time-scale, $T_{\rm m}^{\prime}$, of the NoCool test run and the ratio of the mean merger time-scales, $T_{\rm m}^{\prime}/T_{\rm m}$, are summarized in the top panel.}
\label{fig:inverse_semimajor_axis}
\end{figure*}

\begin{figure} 
\centering\includegraphics[width=\columnwidth]{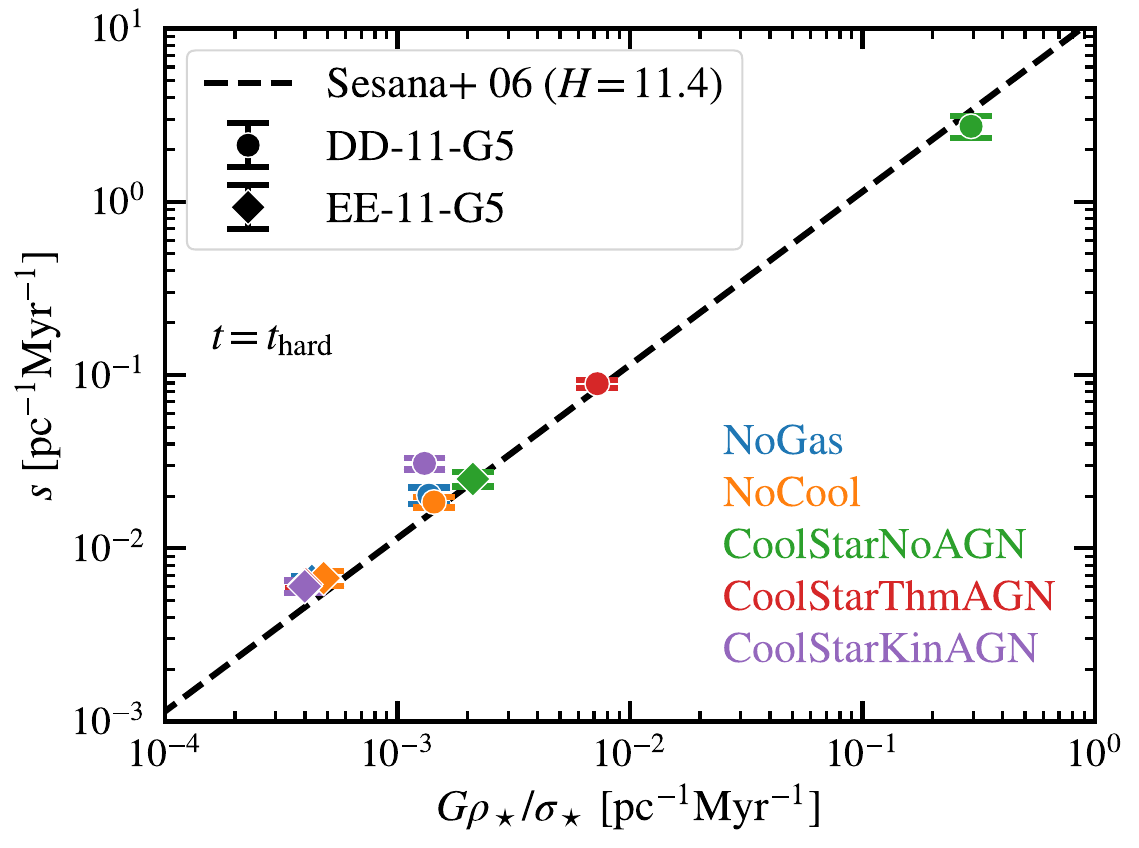}
\caption{Relation between the hardening rate and the stellar properties within the influence radius $r_{\rm m}$ at $t = t_{\rm hard}$. The DD-11-G5 and EE-11-G5 runs are plotted with filled circles and diamonds, respectively. Simulations including different physical processes are distinguished by colours following the same convention used in the other figures. The dashed line shows the hardening rate relation obtained from three-body scattering experiments \citep{Sesana2006}. Note that \citet{Sesana2006} only provide fitting formulae of $H$ for circular binaries with different mass ratios, and thus here $H=11.4$ is computed assuming an eccentricity $e = 0$ and a mass ratio of $q = 1$. According to fig. 3 of \citet{Sesana2006}, at $t_{\rm hard}$, the hardening rate parameters for different eccentricities are fairly close. The agreement between our simulation results and the three-body scattering experiments, which only considers stellar scattering, suggests that the SMBH binary--star interactions are the primary mechanism in driving binary hardening at $t = t_{\rm hard}$ in our simulations.}
\label{fig:hard_rate_Thard}
\end{figure}

\begin{figure*} 
\centering\includegraphics[width=400pt]{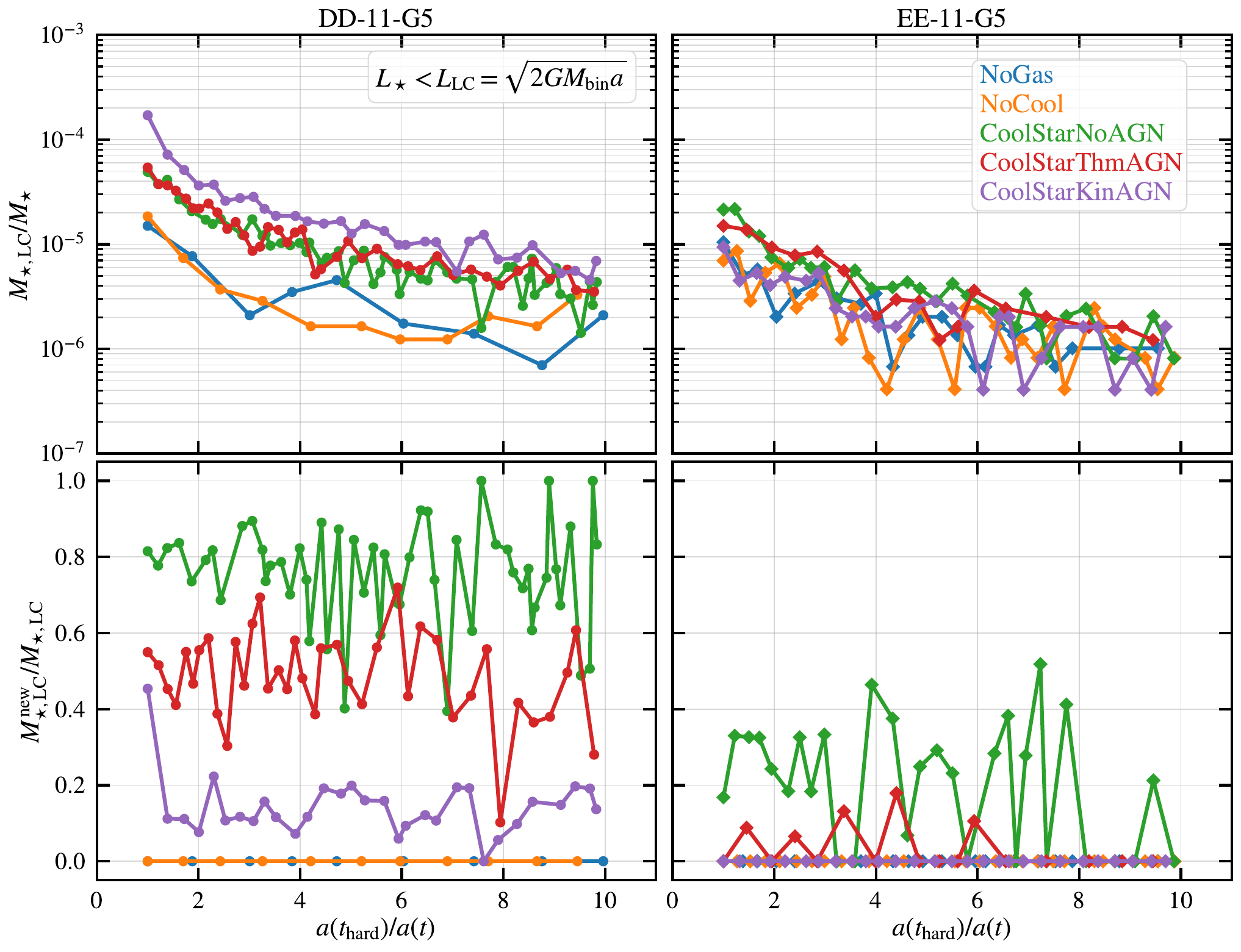}
\caption{Evolution of loss cones for the DD-11-G5 (left) and EE-11-G5 (right) runs. The top panels display the stellar mass fractions within the loss cone, while the bottom panels plot the mass fractions of new stars (i.e. stars formed after $t_0$) in the loss cone. Different simulations are distinguished by colours following the convention used in previous figures. The $x$-axis is the ratio between the semimajor axis at $t_{\rm hard}$ and the semimajor axis at a given time $t$. Only the time period from $t_{\rm hard}$ until the semimajor axis shrinks by a factor of $10$ (before the domination of GW emission in driving the binary hardening) is plotted here.}
\label{fig:loss_cone_evolution}
\end{figure*}

The relations between the dynamical quantities (i.e. the specific energy $\varepsilon$ and specific angular momentum $l$) and the orbital parameters (i.e. $a$ and $e$) of a binary system are given by
\begin{equation}
    \varepsilon = -\frac{GM_{\rm bin}}{2a},
\end{equation}
\begin{equation}
    l^2 = G M_{\rm bin} a (1 - e^2).
\end{equation}
Taking the logarithm and then the time derivative for both sides, we get
\begin{equation}
    \frac{\dot{a}}{a} = \frac{\dot{M}_{\rm bin}}{M_{\rm bin}} - \frac{\dot{\varepsilon}}{\varepsilon},
\end{equation}
\begin{equation}
    \dot{e} = \frac{1 - e^2}{2e}\left[2\frac{\dot{M}_{\rm bin}}{M_{\rm bin}} - \frac{\dot{\varepsilon}}{\varepsilon} - 2 \frac{\dot{l}}{l}\right],
\end{equation}
where the over-dot symbol ($\dot{~}$) denotes the time derivative. If we consider the approximation of $\dot{e} \approx 0$, then
\begin{equation}
    \frac{\dot{a}}{a} = 2 \frac{\dot{l}}{l} - \frac{\dot{M}_{\rm bin}}{M_{\rm bin}}.
\end{equation}
This suggests that the changes in both the total mass of the binary and its specific angular momentum can trigger changes in its orbital evolution. Specifically, SMBH mass accretion leads to a shrinking of the orbit; the decrease in the specific angular momentum due to three-body interactions contributes to the shrinking of the binary orbit; and the change in specific angular momentum due to accretion or torque interactions with surrounding gas (e.g. CBD) will also cause orbital evolution. Furthermore, gas accretion on to SMBHs can introduce additional Brownian motion to the binary and increase the loss cone angular momentum; AGN feedback can alter directly the distribution of surrounding gas and subsequently other galaxy components, thereby impacting the triaxial potential. All of these effects can potentially influence the loss cone refilling rate, and thus affect the binary hardening rate. At the final stage of the binary evolution, the loss of energy and angular momentum due to GW emission dominates and the two SMBHs coalesce rapidly.

Hence, the inclusion of gas and the related galaxy formation processes greatly increase the complexity of the SMBH binary hardening process. In the following subsections, we first analyse the effects of stellar interactions, then study the combined impact from the various galaxy formation processes, and finally assess the influence of the unresolved binary-CBD torque interaction.

\subsection{Effect of stellar interactions}

In Fig.~\ref{fig:inverse_semimajor_axis}, we plot the time-evolution of the inverse semimajor axis, $1/a$, and the hardening rate,
\begin{equation}
    s \equiv \frac{{\rm d}}{{\rm d}t} \left(\frac{1}{a}\right).
\end{equation}
Different simulation sets are distinguished by different colours, and all ten rerun realizations are plotted for each set. Note that the hardening rate is estimated by performing a linear fit to $1/a(t)$ in every interval of $\Delta t = T_{\rm m, max} / N_{\rm bin}$, where $T_{\rm m, max}$ is the maximum merger time-scale across ten realizations and $N_{\rm bin} = 100$. We have performed tests with other values for $N_{\rm bin}$ (e.g. 50 and 200) and confirmed that the presented results are not sensitive to the adopted bin number.

Prior to the dominance of GW emission, which results in a rapid orbital decay, the inverse semimajor axis tends to be a linear function of time (or equivalently the hardening rate tends to be constant). This is particularly noticeable in simulation sets where the three-body interaction is the only mechanism driving SMBH hardening, such as the NoGas and NoCool runs. We also notice that the simulation sets with shorter SMBH merger time-scales generally exhibit higher hardening rates at $t_{\rm hard}$. For instance, in the DD-11-G5 CoolStarNoAGN run, which possesses the shortest $T_{\rm m}$, the hardening rate is $s \sim 3~{\rm pc}^{-1}{\rm Myr}^{-1}$ at $t_{\rm hard}$. On the other hand, in the EE-11-G5 NoCool run, which has the longest $T_{\rm m}$, the hardening rate is much lower, i.e. $s \sim 0.01~{\rm pc}^{-1}{\rm Myr}^{-1}$.

From three-body scattering experiments, it has been shown that the hardening rate is a simple function of the stellar properties \citep[e.g.][]{Quinlan1996,Sesana2006},
\begin{equation}
    s = H\frac{G \rho_{\star}}{\sigma_{\star}},
\end{equation}
where the dimensionless hardening parameter $H$ is approximately a constant once the binary becomes hard, and $\rho_{\star}$ and $\sigma_{\star}$ are the density and velocity dispersion of the stellar background, respectively. 

In Fig.~\ref{fig:hard_rate_Thard}, we compare the relation between $s$ and $G \rho_{\star}/\sigma_{\star}$ at $t_{\rm hard}$ from our reruns to the fitting relation from three-body scattering experiments \citep{Sesana2006}. For each simulation set, the mean and standard deviation of the hardening rate is computed using all ten realizations; the mean density, $\rho_{\star}$, and the one-dimensional velocity dispersion, $\sigma_{\star}$, are computed using the star particles within the influence radius, $r_{\rm m}$. Here, the influence radius is defined as the distance from the binary CoM, within which the enclosed stellar mass is twice the binary mass, i.e. $M_{\star}(<r_{\rm m}) = 2 M_{\rm bin}$.

The relatively good agreement between our simulations and the three-body scattering experiments suggests that the SMBH binary-star interactions are the primary mechanism in driving the binary hardening at $t_{\rm hard}$. Fig.~\ref{fig:hard_rate_Thard} also clearly shows that compared to the case of elliptical galaxy mergers, the more rapid SMBH hardening in disc galaxy mergers is due to the higher central stellar density (or more precisely the higher $\rho_{\star}/\sigma_{\star}$).

In Fig.~\ref{fig:loss_cone_evolution}, we further study the time-evolution of the loss cones during the interval from $t_{\rm hard}$ until the semimajor axis shrinks by a factor of 10 (before entering the GW-driven phase). To achieve better time resolution, we have randomly chosen one realization from each simulation set and increased its snapshot output frequency if necessary.\footnote{Specifically, the original snapshot output interval for all realizations is 50 Myr. In the reruns of the CoolStarNoAGN, CoolStarThmAGN, and CoolStarKinAGN simulation sets of the DD-11-G5 merger with high snapshot cadence, the output intervals are reduced to 0.2, 2, and 5 Myr, respectively. In the EE-11-G5 CoolStarNoAGN rerun, the output interval is reduced to 10 Myr. The remaining simulations sets keep the original output interval.} It is worth noting that other realizations with lower snapshot cadence show qualitatively similar results. The top panels show the fraction of stellar mass enclosed within the loss cone, denoted as $M_{\rm \star, LC}/M_{\star}$, where $M_{\star}$ is the total stellar mass in the simulation and $M_{\rm \star, LC}$ is the total mass of the stars in the loss cone. Here, the loss cone angular momentum is estimated as $L_{\rm LC} = \sqrt{2 G M_{\rm bin} a}$, and a star particle with specific angular momentum\footnote{The specific angular momentum of a star particle is computed with respect to the CoM of the SMBH binary. We have also tested with the CoM of all star particles, and the results are similar.} $L_{\star} < L_{\rm LC}$ is classified as being part of the loss cone population. Overall, the loss cone stellar mass fractions decrease as the binary shrinks and more star particles are ejected from the vicinity of the binary. In disc galaxy mergers, one would expect that the runs including gas cooling and star formation have in general an enhanced fraction of stars in the loss cone. Notably, the CoolStarKinAGN run exhibits the highest mass fractions, primarily attributing to its high $L_{\rm LC}$ -- an outcome of a larger binary mass (arising from the binary becoming hard later and having a longer time to accrete more gas; see Paper II for the detailed SMBH growth history) and a larger hard binary separation (see Fig.~\ref{fig:orb_params_fix_ecc}). In contrast, the mass fractions are more similar across the different simulation sets in the elliptical galaxy mergers due to the smaller differences in star formation and SMBH mass growth.

The bottom panels of Fig.~\ref{fig:loss_cone_evolution} display the mass fractions of new stars (i.e. stars with $t_{\rm SF} > t_0$) in the loss cone. In disc galaxy mergers, ${\sim} 80 \%$ of the stars in the loss cone of the CoolStarNoAGN run are newly formed, followed by the CoolStarThmAGN and CoolStarKinAGN runs with fractions of ${\sim} 50 \%$ and ${\sim} 15 \%$, respectively. For the elliptical galaxy mergers, the CoolStarNoAGN run exhibits a mass fraction of new stars fluctuating around ${\sim} 20\%$ and the fraction reduces as the binary hardens; the CoolStarThmAGN only shows non-zero fractions in a few snapshots. This comparison highlights the crucial role of new stars in replenishing the loss cone and accelerating the process of binary hardening, particularly for gas-rich disc galaxy mergers.

The analyses in this subsection point to the following two step picture in relation to SMBH binary hardening following a galaxy merger: firstly, due to radiative cooling and tidal torques, gas condenses to the galaxy centre and forms stars, boosting the central stellar density. Secondly, these recently formed stars -- inheriting the low angular momenta from the gas -- contribute to the loss cone  and assist in the SMBH hardening process via three-body interactions. This process is particularly important in gas-rich disc galaxy mergers, resulting in shorter SMBH merger time-scales, compared to the gas-poor elliptical galaxy mergers. The comparison across different simulation sets highlights the importance of modelling galaxy formation processes, which affect gas cooling and star formation, in studying the SMBH coalescence process.

It is worth noting that \citet{Liao2023} already demonstrated a tendency for the stellar particles ejected by the SMBH binary during the hardening phase to have young ages, thereby providing additional support for the picture described above. The simulation results from \citet{Khan2016} also strengthen this picture. By extracting a galaxy merger remnant at $z \sim 3$ from a cosmological simulation and further evolving an SMBH binary in the remnant centre with the direct $N$-body code {\sc $\phi$-gpu}, \citet{Khan2016} found that due to the high central stellar density, which is attributed to recent star formation, the SMBH merger time-scale is as short as ${\sim} 10$ Myr. Intriguingly, in the absence of AGN feedback in both their parent cosmological simulation and zoomed simulations, their SMBH merger time-scale is quite close to that of our DD-11-G5 CoolStarNoAGN run (i.e. ${\sim} 10$ Myr versus ${\sim} 20$ Myr). However, our results suggest that once AGN feedback is included, the SMBH merger time-scales become much longer at ${\sim} 100$--$500$ Myr.

\subsection{Combined effects of galaxy formation processes}

During the hardening phase, as the SMBH binary is dynamically ejecting stars via the gravitational slingshot interaction, the various galaxy formation processes also come into play. Specifically, SMBHs accrete gas and release feedback energy, which increases the SMBH masses, introduces Brownian motions, and alters the surrounding matter distribution, directly affecting the binary hardening. Simultaneously, gas cools and turns into stars stochastically (if the star formation conditions are met) and stars influence surrounding gas via stellar feedback processes. These combined effects can further impact the distribution of matter and the repopulation of the loss cone. Given that all of these processes occur simultaneously and interplay with each other, it is challenging to distinctly isolate each effect.

To study the combined impact of these galaxy formation processes on the SMBH coalescence, we perform the `NoCool test' reruns -- starting from the same initial conditions as the CoolStarNoAGN, CoolStarThmAGN, and CoolStarKinAGN reruns but running them with the NoCool model, which only considers gravity and SPH with an adiabatic equation of state. These NoCool test reruns (ten realizations for each set) are plotted with a brown colour in Fig.~\ref{fig:inverse_semimajor_axis}. In the corresponding `Info' panels, we summarize the SMBH merger time-scales for both the original reruns ($T_{\rm m}$) and the NoCool test reruns ($T_{\rm m}^\prime$), including both the mean and the standard deviation. In addition, we calculate the ratio of their mean merger time-scales ($T_{\rm m}^\prime/T_{\rm m}$). Moreover, we provide the averaged star formation rate (SFR) over $T_{\rm m}$ and the averaged SMBH mass growth factor ($M_{\rm bin, f}/M_{\rm bin, i}$ with $M_{\rm bin, i}$ being the binary mass at $t_{\rm hard}$ and $M_{\rm bin, f}$ being the binary mass at $t_{\rm hard} + T_{\rm m}$) for the original reruns.

By comparing the NoCool test reruns with the original reruns, we can obtain an estimation of the impact from the galaxy formation processes.\footnote{Strictly speaking, extracting a snapshot from a simulation including radiative cooling and feedback processes, and subsequently rerunning it using the NoCool model, introduces a rapid change in the gas distribution after the start of the simulation. This dynamic change in the gravitational potential could also contribute to the differences in the SMBH evolution. Nevertheless, despite these intricacies, the NoCool tests can still provide valuable insight into the impact from the galaxy formation processes.} Interestingly, all runs except DD-11-G5 CoolStarNoAGN and EE-11-G5 CoolStarKinAGN have an SMBH merger time-scale ratio of $T_{\rm m}^\prime/T_{\rm m} \sim 1.7$, indicating that, on average, the inclusion of galaxy formation processes has shortened the SMBH merger time-scales by a factor of ${\sim} 1.7$. In the case of DD-11-G5 CoolStarNoAGN, the high hardening rate due to the existing stellar background overshadows the impact of gas cooling and star formation on the SMBH merger time-scale. In the case of EE-11-G5 CoolStarKinAGN, the low star formation rate (${\sim} 0.01~{\rm M}_{\sun}{\rm yr}^{-1}$) and minimal SMBH mass growth (${\sim}1\%$) due to the effective kinetic AGN feedback result in a smaller reduction in the merger time-scale (only by a factor of ${\sim} 1.1$).

\subsection{Effects from the unresolved binary-CBD torque interactions}

\begin{figure*} 
\centering\includegraphics[width=500pt]{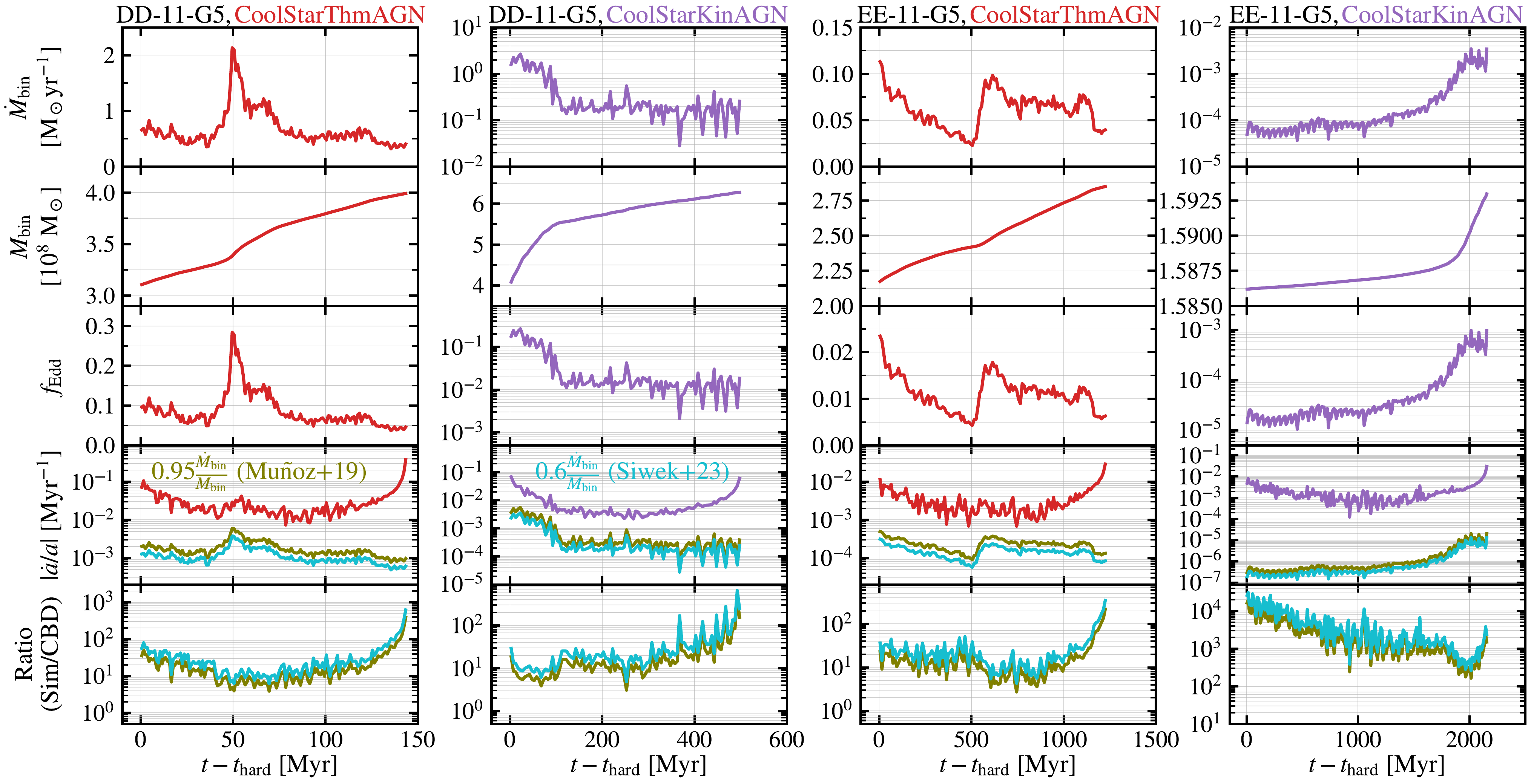}
\caption{Comparison between our simulations and the expected effects from the CBD torques. The two left columns show the results from the CoolStarThmAGN (red) and CoolStarKinAGN (purple) runs of the DD-11-G5 merger, while the two right columns plot those runs of the EE-11-G5 merger. From top to bottom, we plot the total SMBH binary accretion rate ($\dot{M}_{\rm bin}$), the SMBH binary mass ($M_{\rm bin}$), the Eddington ratio ($f_{\rm Edd} \equiv \dot{M}_{\rm bin} / \dot{M}_{\rm Edd}$), the magnitude of the SMBH binary inspiral/outspiral rate ($|\dot{a}/a|$), and the ratio between the inspiral rate from our simulations and the outspiral rate expected from the CBD torques. In the bottom two rows, the olive and cyan curves show the results of the outspiral SMBH binary from CBD simulations of \citet{Munoz2019} and \citet{Siwek2023}, respectively. Note that the plotted physical quantities have been averaged over 1 Myr, 5 Myr, 10 Myr, and 10 Myr for the runs from left to right.}
\label{fig:hardening_rate_cbd}
\end{figure*}

In our subgrid model for binary SMBH accretion, we incorporate the preferential accretion mechanism motivated by the small-scale hydrodynamical CBD simulations, i.e. the secondary SMBH tends to have a higher accretion rate than the primary SMBH. However, our SMBH subgrid model does not include the binary--CBD torque interaction, which can also affect the orbital evolution of the binary. This omission stems partly from the ongoing debates within this field and the large uncertainty in the existing results. To elaborate, early CBD literature suggested a negative gravitational torque from the CBD acting upon the binary, which caused the binary's orbit to shrink \citep[e.g.][]{Artymowicz1991,MacFadyen2008}. On the contrary, some recent CBD simulations have proposed a positive torque: while the torque from the outer CBD is negative, as previously suggested, the torque from the gas within the inner streams and mini-discs can be overwhelmingly positive. This results in a net positive torque and, consequently, the expansion of the binary \citep[e.g.][]{Miranda2017,Tang2017,Moody2019,Munoz2019,Munoz2020,Duffell2020}. Subsequent simulation studies have demonstrated that whether an SMBH binary shrinks or expands can depend on various factors, including the specific disc parameters such as the disc thickness and viscosity \citep[e.g.][]{Heath2020,Tiede2020,Dittmann2022,Dittmann2023}, the binary properties like the SMBH mass ratio and binary eccentricity \citep[e.g.][]{DOrazio2021,Siwek2023}, and the detailed simulation setups (e.g. 2D versus 3D, inclusion of CBD self-gravity or not, and fixed versus live SMBH binaries; see e.g. \citealt{Franchini2021,Franchini2022,Franchini2023}).

In the following, we estimate the impact of the binary--CBD torque interaction on our binary hardening results. To achieve this, we compare the binary inspiral/outspiral rate, $\dot{a}/a$, measured from CBD simulations with the inspiral rate directly measured from our simulations.

From CBD simulations, it has been found that under the binary--CBD interaction, an eccentric and equal-mass binary with $e \ga 0.1$ evolves towards an equilibrium eccentricity of $e_{\rm eq} \sim 0.5$ while a nearly circular and equal-mass binary ($e \la 0.1$) stays in a circular orbit with $e_{\rm eq} \sim 0$ \citep[e.g.][]{DOrazio2021,Zrake2021,Siwek2023}. This implies that our simulated binary, which has $M_{\rm BH,2}/M_{\rm BH,1} \approx 1$ and $e_{\rm hard} = 0.6$, will settle at $e \approx 0.5$ within a few Myr \citep{Siwek2023} when the binary--CBD interaction is considered.

In the CBD study, the inspiral/outspiral rate of the binary orbit is usually expressed using the specific accretion rate of the binary system. For an equal-mass binary with $e = 0.5$, \citet{Munoz2019} provided the following fitting relation from their simulations (see their table 1),
\begin{equation}\label{eq:munoz2019}
    \frac{\dot{a}}{a} = 0.95 \frac{\dot{M}_{\rm bin}}{M_{\rm bin}};
\end{equation}
\citet{Siwek2023} gave a similar relation (see their figure 2),
\begin{equation}\label{eq:siwek2023}
    \frac{\dot{a}}{a} = 0.6 \frac{\dot{M}_{\rm bin}}{M_{\rm bin}}.
\end{equation}

Note that in both relations, $\dot{a}/a$ is positive, implying that the binary expands due to the torque interaction with its CBD. It is important to compare the outspiral rate from CBD simulations, which consider only the binary--CBD interaction, with the inspiral rate from our simulations, which consider the binary--star interaction and the effects from other galaxy formation processes, to see which mechanism dominates and whether our simulated binary shrinks or expands.

This comparison is summarized in Fig.~\ref{fig:hardening_rate_cbd} for the reruns with AGN feedback. Note that different rerun realizations of the same simulation set exhibit qualitatively similar results, therefore only one representative realization from each set is displayed here. The upper three rows present the time-evolution of the binary accretion rate, the binary mass, and the Eddington ratio of the binary system. All runs show binary accretion rates below the Eddington limit, with $f_{\rm Edd} \la 0.3$. The fourth row displays the magnitude of the inspiral rate (i.e. $-\dot{a}/a$) from our simulations (in red) and the estimated CBD-induced outspiral rates according to Eqs (\ref{eq:munoz2019}) and (\ref{eq:siwek2023}) (in olive and cyan respectively). It is evident that in all runs, the inspiral rates attributed to the binary--star interaction and other galaxy formation processes are higher than the outspiral rates resulting from the binary--CBD torque interaction. According to the bottom panel, the former is at least a few times higher than the latter. For runs with low accretion rates (e.g. EE-11-G5 CoolStarKinAGN), the ratio is notably within the range of ${\sim} 10^{2}$--$10^{4}$.

This quantitative comparison suggests that for our galaxy mergers, prior to the GW emission-dominated phase, the primary driving mechanisms for binary orbital evolution are the binary--star slingshot interaction and other galaxy formation processes. The binary--CBD torque interaction appears to play only a complementary and secondary role. In gas-rich mergers, despite the high binary accretion and CBD-induced outspiral rates, efficient star formation boosts the central stellar density and the inspiral rate. Conversely, gas-poor mergers exhibit low binary accretion rates, resulting in even lower CBD-induced outspiral rates despite the low inspiral rates from binary-star interactions. Our results overall agree with the work of \citet{Bortolas2021}, which suggests that CBDs have a minimal impact on binary evolution when accretion rates are significantly lower than the Eddington limit.

\section{Conclusions} \label{sec:con}

In this work, we utilized the simulations of the RABBITS series to study how various galaxy formation processes (e.g. cooling, star formation, SMBH accretion, stellar and AGN feedback) influence the orbital evolution of SMBH binaries during the hardening and GW emission phases. Especially, we systematically mitigated the impact of the stochastic binary eccentricity at $t_{\rm hard}$ by adjusting it to the same value across the different simulations.

Our major findings are summarized as follows.

\begin{enumerate}
    \item Gas-rich disc galaxy mergers exhibit significantly shorter SMBH merger time-scales than gas-poor elliptical mergers. For eccentricities of $e \sim 0.6$--$0.8$ during the hardening phase, the simulations with AGN feedback reveal merger time-scales of $T_{\rm m} \sim 100$--$500$ Myr for disc mergers and $T_{\rm m} \sim 1$--$2$ Gyr for elliptical mergers. Elliptical merger runs with kinetic AGN feedback yield similar time-scales to gas-free runs, supporting the gas-free assumption for simulating elliptical galaxy mergers adopted in previous studies.
    \item Nuclear (or centrally concentrated) star formation has a strong influence on SMBH merger time-scales. Throughout the galaxy merging process, the gas condensation at the galaxy centre due to radiative cooling and tidal torques leads to star formation in the nuclear region of the galaxy, boosting the central stellar density (i.e. within the radius of a few hundred parsecs). These recently formed stars, inheriting low angular momenta from the gas, contribute to the loss cone, assisting binary hardening, especially in gas-rich disc mergers. This also explains why the simulations with cooling and star formation but without AGN feedback (i.e. the CoolStarNoAGN runs) exhibit the shortest merger time-scales.
    \item After an SMBH binary becomes hard, compared to non-radiative hydrodynamical runs (i.e. the NoCool test runs), galaxy formation processes collectively shorten the SMBH merger time-scale by a factor of ${\sim} 1.7$ on average.
    \item In our simulated retrograde galaxy mergers, the orbital evolution of SMBH binaries is primarily driven by binary--star gravitational slingshot interactions and other galaxy formation processes, with the binary--CBD torque interaction only playing a minor role. The outspiral evolution due to the binary--CBD torque interaction has minimal impact on the shrinking of the orbit and SMBH merger time-scales in our studied mergers.
\end{enumerate}

Our study highlights the crucial role of nuclear star formation in driving the coalescence of SMBHs, especially in gas-rich disc galaxy mergers. It also reinforces the importance of including galaxy formation processes in improving the predictions of SMBH merger time-scales.

In the current work, we have considered only the G5 galaxy merger geometry and fixed $e_{\rm hard}$ to a single value of 0.6. To further explore the parameter space and investigate the relative roles of the binary--star and binary--CBD interactions, in our future RABBITS studies, we will consider other merger geometries \citep[see e.g.][]{Naab2003}, which can result in different strengths of tidal response and nuclear starburst, and adjust $e_{\rm hard}$ to other values. These upcoming investigations are expected to further deepen our understanding in the hardening and coalescing process of SMBH binaries, a key target of low-frequency GW observations.

\section*{Acknowledgements}
We thank the anonymous referee for helpful comments. SL, DI, PHJ, FPR, and RJW acknowledge the support by the European Research Council via ERC Consolidator Grant KETJU (no. 818930). PHJ and JMH acknowledge the support by the Academy of Finland grant no. 339127. SL also acknowledges the supports by the National Natural Science Foundation of China (NSFC) grant (No. 11988101) and the K. C. Wong Education Foundation. TN acknowledges support from the Deutsche Forschungsgemeinschaft (DFG, German Research Foundation) under Germany’s Excellence Strategy -- EXC-2094 -- 390783311 from the DFG Cluster of Excellence ``ORIGINS''. AR acknowledges the support of the University of Helsinki Research Foundation. The numerical simulations used computational resources provided by the CSC -- IT Center for Science, Finland. This research has made use of NASA’s Astrophysics Data System. We gratefully thank the developers of the open-source \textsc{python} packages that were used in the data analysis of this work, including \textsc{matplotlib} \citep{Hunter2007}, \textsc{numpy} \citep{Harris2020}, \textsc{scipy} \citep{Virtanen2020}, \textsc{astropy} \citep{astropy2013,astropy2018,astropy2022}, and \textsc{pygad} \citep{Rottgers2020}.

\section*{Data availability}
The simulation data used in this article will be shared upon a reasonable request to the corresponding author.

\bibliographystyle{mnras}
\bibliography{ref} 

\appendix

\section{Resetting SMBH binary eccentricities}\label{ap:reset_ecc}

For the two SMBHs in a hard binary, we assume their coordinates are denoted as $\mathbfit{r}_1$ and $\mathbfit{r}_2$, their velocities $\mathbfit{v}_1$ and $\mathbfit{v}_2$, and their masses $m_1$ and $m_2$. Then, their CoM position and velocity are
\begin{equation}
    \mathbfit{r}_{\rm CoM} = \frac{m_1 \mathbfit{r}_1 + m_2 \mathbfit{r}_2}{m_1 + m_2}, \mathbfit{v}_{\rm CoM} = \frac{m_1 \mathbfit{v}_1 + m_2 \mathbfit{v}_2}{m_1 + m_2},
\end{equation}
respectively. The semimajor axis and the eccentricity of the binary are
\begin{equation}
    a = -\frac{\mu}{2\varepsilon}, e = \sqrt{1 + \frac{2\varepsilon h^2}{\mu^2}},
\end{equation}
respectively. Here, the standard gravitational parameter is
\begin{equation}
    \mu = G (m_1 + m_2),
\end{equation} 
the specific orbital energy is
\begin{equation}
    \varepsilon = \frac{v_{\rm rel}^2}{2} - \frac{\mu}{r_{\rm rel}},
\end{equation}
the specific relative angular momentum is
\begin{equation}
    h = |\mathbfit{h}| = |\mathbfit{r}_{\rm rel} \times \mathbfit{v}_{\rm rel}|,
\end{equation}
and the relative position and the relative velocity are
\begin{equation}
    r_{\rm rel} = |\mathbfit{r}_{\rm rel}| = |\mathbfit{r}_1 - \mathbfit{r}_2|, v_{\rm rel} = |\mathbfit{v}_{\rm rel}| = |\mathbfit{v}_1 - \mathbfit{v}_2|.
\end{equation}
To reset the eccentricity of this binary to any given values in the range of $0 \leq e_{\rm new} < 1$, we will keep $m_1$, $m_2$, $\mathbfit{r}_{\rm CoM}$, $\mathbfit{v}_{\rm CoM}$, and $a$ fixed, and change the coordinates and velocities of the two SMBHs, i.e. $\mathbfit{r}_{\rm 1, new}$, $\mathbfit{r}_{\rm 2, new}$, $\mathbfit{v}_{\rm 1, new}$, and $\mathbfit{v}_{\rm 2, new}$.

For a given $e_{\rm new}$, the new specific relative angular momentum is
\begin{equation} \label{eq:ecc_h}
    h_{\rm new} = r_{\rm rel, new} v_{\rm rel, new} \sin \delta = \sqrt{\frac{\mu^2 (e_{\rm new}^2 - 1)}{2\varepsilon}},
\end{equation}
where $r_{\rm rel, new} = |\mathbfit{r}_{\rm rel, new}| = |\mathbfit{r}_{\rm 1, new} - \mathbfit{r}_{\rm 2, new}|$, $v_{\rm rel, new} = |\mathbfit{v}_{\rm rel, new}| = |\mathbfit{v}_{\rm 1, new} - \mathbfit{v}_{\rm 2, new}|$, and $\delta$ is the angle between $\mathbfit{r}_{\rm rel, new}$ and $\mathbfit{v}_{\rm rel, new}$. The new distance and the new relative velocity magnitude can be set to 
\begin{equation}
    r_{\rm rel, new} = -\frac{\mu}{2\varepsilon},  v_{\rm rel, new} = \sqrt{-2 \varepsilon},
\end{equation}
respectively, so that for any $e_{\rm new}$ in the range $[0,1)$, the solution of $\delta$ exists. Note that Eq. (\ref{eq:ecc_h}) only specifies the magnitude of the specific relative angular momentum, we can set different directions for $\mathbfit{h}_{\rm new}$, which is achieved by adjusting the directions of $\mathbfit{r}_{\rm rel, new}$ and $\mathbfit{v}_{\rm rel, new}$ as detailed in the following, to create different realizations.

The new coordinates for the two SMBHs can be set to
\begin{equation}
    \mathbfit{r}_{\rm 1, new} = \mathbfit{r}_{\rm CoM} + \frac{m_2}{m_1 + m_2} \mathbfit{r}_{\rm rel, new},
\end{equation}
\begin{equation}
    \mathbfit{r}_{\rm 2, new} = \mathbfit{r}_{\rm CoM} - \frac{m_1}{m_1 + m_2} \mathbfit{r}_{\rm rel, new},
\end{equation}
where $\mathbfit{r}_{\rm rel, new} = r_{\rm rel, new} \hat{\mathbfit{r}}$ with $\hat{\mathbfit{r}}$ being a unit vector with uniformly distributed random directions.

To fully determine $\mathbfit{v}_{\rm rel, new}$, apart from the magnitude $v_{\rm rel, new}$ and the angle $\delta$, we still need another angle $\eta$. Here $\delta$ and $\eta$ can be regarded as the polar angle and the azimuthal angle in a spherical coordinate system $S^\prime$ with $\hat{\mathbfit{r}}$ being the $z$-direction. The angle $\eta \in [0, 2\pi)$ can be treated as a free parameter and drawn from a uniform distribution. In the spherical coordinate system $S^\prime$, the new relative velocity is 
\begin{equation}
    \mathbfit{v}_{\rm rel, new}^\prime = (v_{\rm rel, new} \sin \delta \cos \eta, v_{\rm rel, new} \sin \delta \sin \eta, v_{\rm rel, new} \cos \delta).
\end{equation}
To obtain the expression of $\mathbfit{v}_{\rm rel, new}$ in the original simulation coordinate system $S$, we need to specify the spherical coordinate system $S^\prime$ and the rotation matrix between $S^\prime$ and $S$. This can be done using the Euler angles $(\alpha, \beta, \gamma)$. Here we adopt the $z$-$x$-$z$ proper Euler angle convention, i.e. $\alpha$ is the rotation angle around the $z$ axis, $\beta$ is the rotation angle around the $x^\prime$ axis, and $\gamma$ is the rotation angle around the $z^{\prime\prime}$ axis. As a simple choice, we adopt $\gamma = 0$. Then, the relation between $\mathbfit{v}_{\rm rel, new}$ and $\mathbfit{v}_{\rm rel, new}^\prime$ is
\begin{equation}
    \mathbfit{v}_{\rm rel, new} = R \mathbfit{v}_{\rm rel, new}^\prime,
\end{equation}
where the rotation matrix is
\begin{equation}
    R = 
    \begin{bmatrix}
\cos \alpha & -\sin \alpha \cos \beta & \sin \alpha \sin \beta\\
\sin \alpha & \cos \alpha \cos \beta & -\cos \alpha \sin \beta \\
0 & \sin \beta & \cos \beta
\end{bmatrix},
\end{equation}
and
\begin{equation}
    \sin \alpha = \frac{x}{\sqrt{x^2 + y^2}}, ~ \cos \alpha = -\frac{y}{\sqrt{x^2 + y^2}},
\end{equation}
\begin{equation}
    \sin \beta = \frac{\sqrt{x^2 + y^2}}{r_{\rm rel, new}}, ~ \cos \beta = \frac{z}{r_{\rm rel, new}}.
\end{equation}
Here, $x$, $y$, and $z$ are the three components of $\mathbfit{r}_{\rm rel, new}$ in the original simulation coordinate system $S$, i.e. $\mathbfit{r}_{\rm rel, new} \equiv (x, y, z)$.

Finally, the new velocities for the two SMBHs are
\begin{equation}
    \mathbfit{v}_{\rm 1, new} = \mathbfit{v}_{\rm CoM} + \frac{m_2}{m_1 + m_2} \mathbfit{v}_{\rm rel, new},
\end{equation}
\begin{equation}
    \mathbfit{v}_{\rm 2, new} = \mathbfit{v}_{\rm CoM} - \frac{m_1}{m_1 + m_2} \mathbfit{v}_{\rm rel, new}.
\end{equation}

\section{Resolution study} \label{ap:res_study}

\begin{table*}
\begin{threeparttable}
\caption{Details of the CoolStarKinAGN simulations with different numerical resolutions. From the left, the galaxy merger, the simulation name, the number of star particles at $t_{\rm hard}$ ($N_{\star}$), the number of gas particles at $t_{\rm hard}$ ($N_{\rm gas}$), the number of dark matter particles ($N_{\rm DM}$), the averaged star/gas particle mass ($m_{\star}$ or $m_{\rm gas}$), the dark matter particle mass ($m_{\rm DM}$), the stellar softening length ($\epsilon_{\star}$), the gas softening length ($\epsilon_{\rm gas}$), the dark matter softening length ($\epsilon_{\rm DM}$), and the mean and standard deviation of the SMBH merger time-scale ($T_{\rm m}$).}
\label{tab:ap_sim_info}
\begin{tabular}{llccccccccccccc}
\hline
Galaxy & Simulation & $N_{\star}$ & $N_{\rm gas}$ & $N_{\rm DM}$ & $m_{\star}$ and $m_{\rm gas}$ & $m_{\rm DM}$ & $\epsilon_{\star}$ & $\epsilon_{\rm gas}$ & $\epsilon_{\rm DM}$ & $T_{\rm m}$\\
merger & name & [$10^{6}$] & [$10^{6}$] & [$10^{6}$] & [$10^{5}~{\rm M}_{\sun}$] & [$10^{5}~{\rm M}_{\sun}$] & [pc] & [pc] & [pc] & [Myr]\\
\hline
DD-11-G5 & Low-res-$N_{\rm p}/4$ & 0.65 & 0.07 & 0.80 & 4.0 & 62.9 & 8.0 & 32.0 & 160.0 & 665.3 $\pm$ 156.7 \\
DD-11-G5 & Low-res-$N_{\rm p}/2$ & 1.32 & 0.12 & 1.60 & 2.0 & 31.0 & 6.0 & 25.0 & 126.0 & 563.6 $\pm$ 64.4  \\
DD-11-G5 & Fiducial-$N_{\rm p}$ & 2.68 & 0.19 & 3.20 & 1.0 & 15.5 & 5.0 & 20.0 & 100.0 & 453.8 $\pm$ 77.5 \\
DD-11-G5 & Hi-res-$2N_{\rm p}$ & 5.44 & 0.32 & 6.40 & 0.5 & 7.7 & 4.0 & 16.0 & 80.0 & 402.8 $\pm$ 40.0 \\
\\
EE-11-G5 & Low-res-$N_{\rm p}/4$ & 0.61 & 0.13 & 0.80 & 4.0 & 62.9 & 8.0 & 32.0 & 160.0 & 2363.1 $\pm$ 565.9 \\
EE-11-G5 & Low-res-$N_{\rm p}/2$ & 1.22 & 0.26 & 1.60 & 2.0 & 31.0 & 6.0 & 25.0 & 126.0 & 2377.8 $\pm$ 548.9 \\
EE-11-G5 & Fiducial-$N_{\rm p}$ & 2.44 & 0.53 & 3.20 & 1.0 & 15.5 & 5.0 & 20.0 & 100.0 & 2107.2 $\pm$ 424.1 \\
EE-11-G5 & Hi-res-$2N_{\rm p}$ & 4.87 & 1.06 & 6.40 & 0.5 & 7.7 & 4.0 & 16.0 & 80.0 & 2331.5 $\pm$ 213.1 \\
\hline
\end{tabular}
\end{threeparttable}
\end{table*}

\begin{figure*} 
\centering\includegraphics[width=400pt]{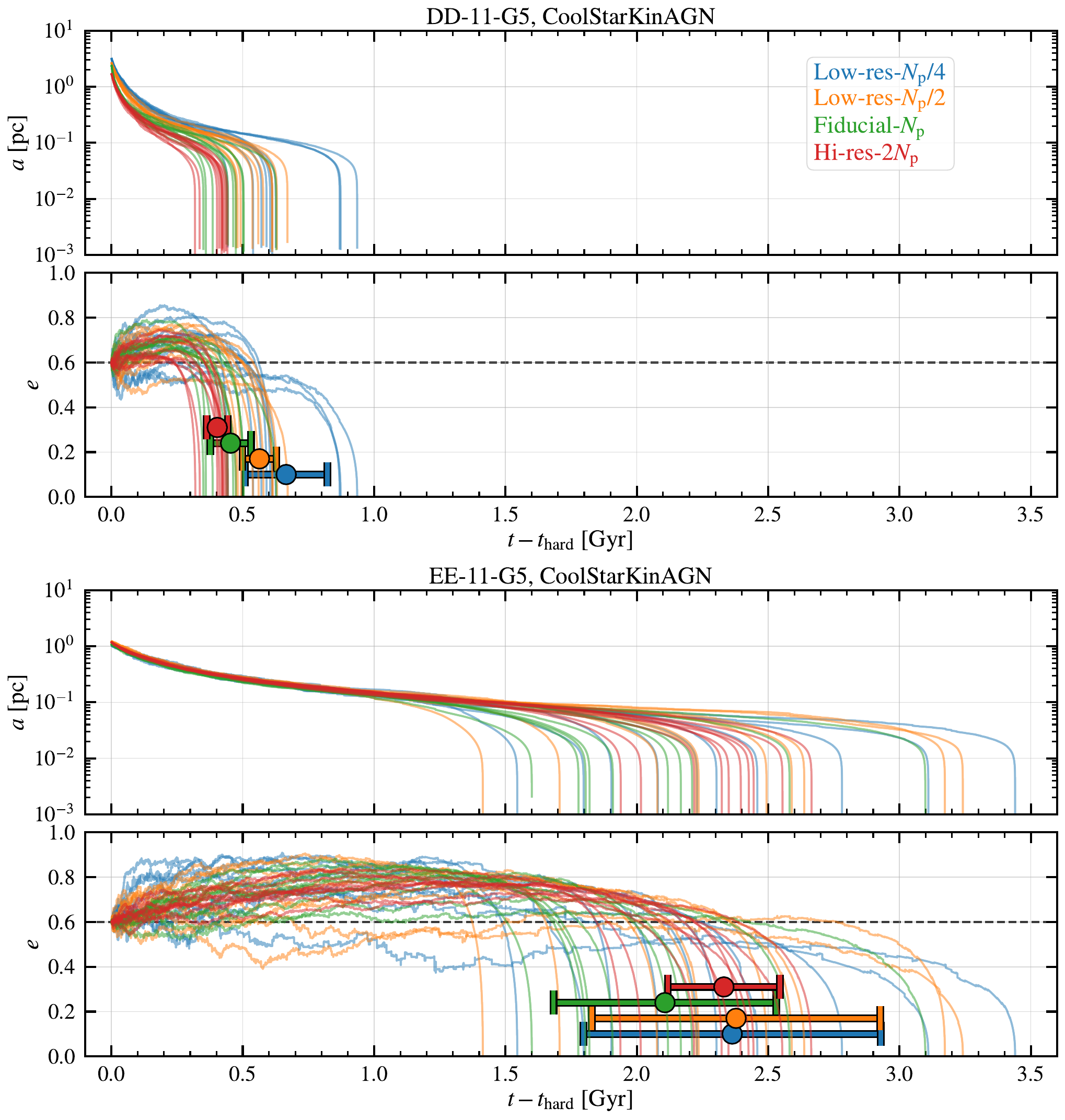}
\caption{Similar as Fig.~\ref{fig:orb_params_fix_ecc}, but for the CoolStarKinAGN resolution study runs. The Low-res-$N_{\rm p}/4$, Low-res-$N_{\rm p}/2$, Fiducial-$N_{\rm p}$, and Hi-res-$2N_{\rm p}$ sets are plotted with the blue, orange, green, and red colours, respectively. For each resolution set, the filled circle and the error bar show the mean SMBH coalescence time over ten realizations and their standard deviation. The DD-11-G5 test runs show a weak resolution dependence for the SMBH merger time-scales, while the merger time-scales in the EE-11-G5 runs do not exhibit resolution dependences; see Fig.~\ref{fig:merger_timescales_res_study} for details.}
\label{fig:orb_params_fix_ecc_res_study}
\end{figure*}

\begin{figure} 
\centering\includegraphics[width=\columnwidth]{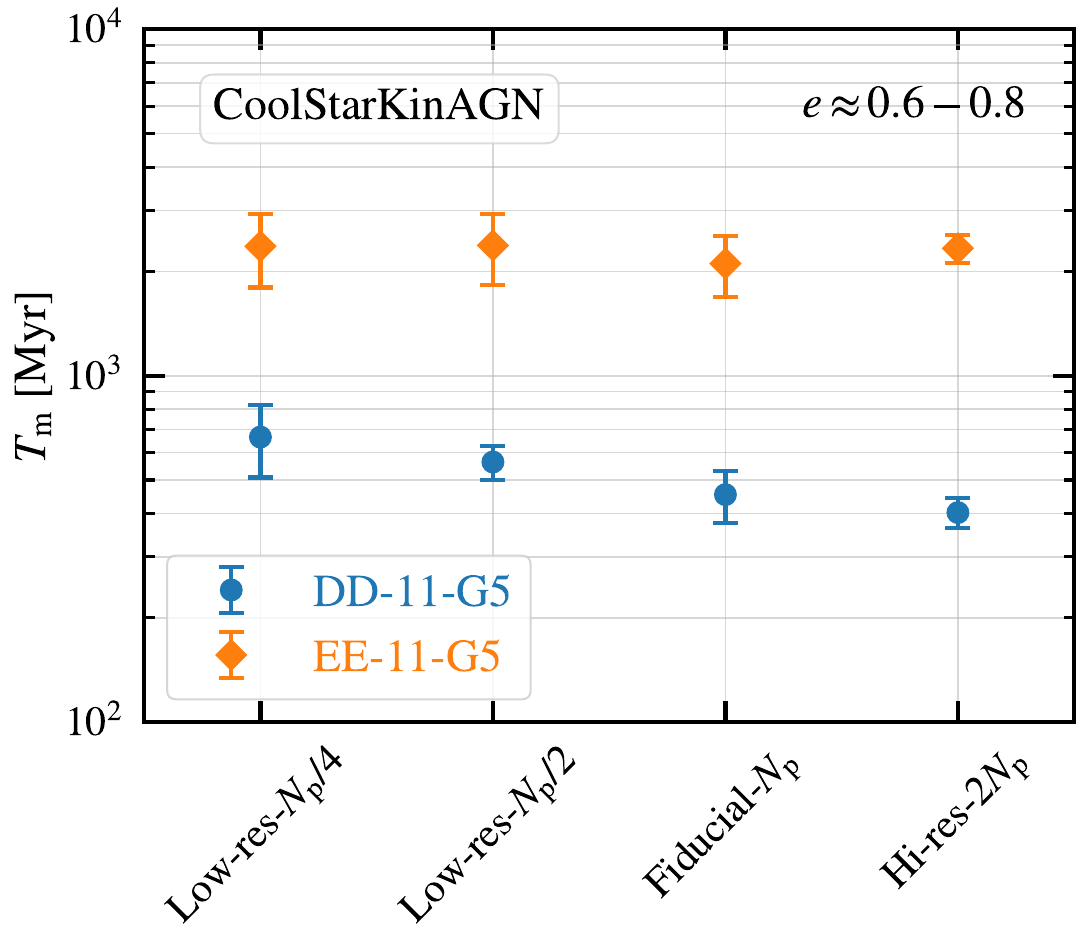}
\caption{Similar as Fig.~\ref{fig:merger_timescales}, but for the CoolStarKinAGN resolution test runs. For higher resolution runs, the standard deviation of the SMBH merger time-scales tends to be smaller. In the DD-11-G5 runs, the merger time-scales show a weak resolution dependence (i.e. with shorter $T_{\rm m}$ for runs with higher resolutions), but the merger time-scale from the Fiducial-$N_{\rm p}$ runs agrees well with that from the Hi-res-$2N_{\rm p}$ runs within the one-sigma scatter. In the EE-11-G5 simulations, there is no clear resolution dependence for the merger time-scales.}
\label{fig:merger_timescales_res_study}
\end{figure}

\begin{figure} 
\centering\includegraphics[width=\columnwidth]{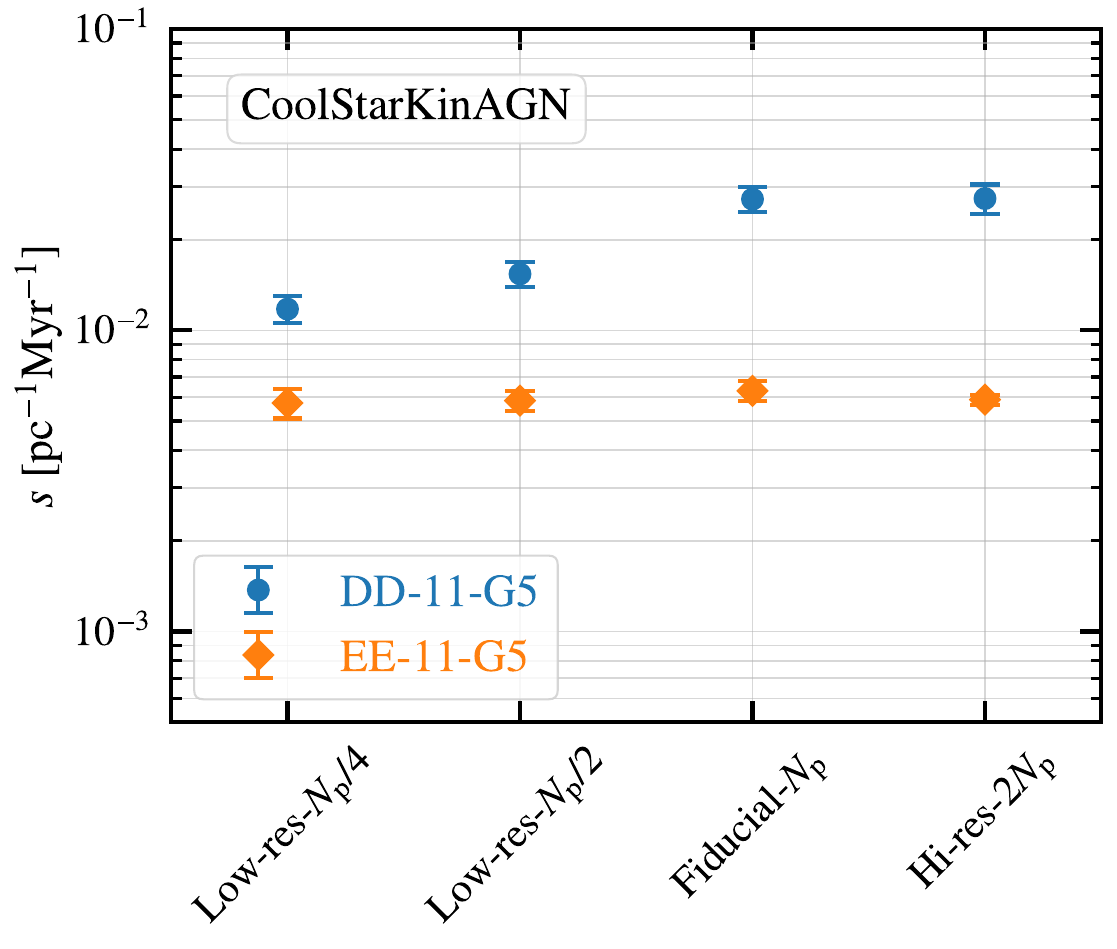}
\caption{Binary hardening rates from the CoolStarKinAGN reruns with different particle numbers. The hardening rates here are estimated by performing linear fits to $1/a(t)$ in the interval $\Delta t$ from $t_{\rm hard}$ to the time when $a(t) = a(t_{\rm hard})/2$. The DD-11-G5 and EE-11-G5 results are shown with blue and orange colours, respectively. The mean and standard deviation computed from the ten realizations of each resolution set are plotted with the filled circles/diamonds and error bars. For the DD-11-G5 runs, the hardening rates show a dependence on particle number in the low-resolution runs, however, the Fiducial-$N_{\rm p}$ run converges well to the Hi-res-$2N_{\rm p}$ run. For the EE-11-G5 runs, there is no dependence of hardening rates on particle number.}
\label{fig:hardening_rate_res_study}
\end{figure}

In this appendix, we conduct a resolution convergence test to assess the impact of numerical resolution on the SMBH merger time-scales. Similar to the fiducial CoolStarKinAGN reruns introduced in the main text, we adopt the snapshots at $t_{\rm hard}$ from the parent runs with different resolutions, which are presented in appendix A of Paper II. The binary eccentricity is set to $e_{\rm hard} = 0.6$, generating ten realizations for each parent run. These reruns are then evolved until the SMBH merger occurs.

In Table~\ref{tab:ap_sim_info}, we summarize the particle numbers, masses, and softening lengths for the reruns. The fiducial CoolStarKinAGN reruns discussed in the main text are termed `Fiducial-$N_{\rm p}$'. Simulations with mass resolutions lowered by a factor of 2 and 4 are called `Low-res-$N_{\rm p}/2$' and `Low-res-$N_{\rm p}/4$', respectively. The simulation with higher mass resolution (i.e. a factor of 2 higher than the fiducial one) is denoted as `Hi-res-$2N_{\rm p}$'.

The time-evolution of the orbital parameters from all reruns are displayed in Fig.~\ref{fig:orb_params_fix_ecc_res_study}. A discernible trend is observed: as the mass resolution increases, the scatter among the realizations is reduced. The SMBH merger time-scales are summarized in the rightmost column of Table~\ref{tab:ap_sim_info} and also shown in Fig.~\ref{fig:merger_timescales_res_study}. In the DD-11-G5 reruns, the merger time-scales exhibit a weak resolution dependence, with shorter time-scales for higher resolutions. However, the merger time-scale from the Fiducial-$N_{\rm p}$ reruns agrees well with that from the Hi-res-$2N_{\rm p}$ reruns within the $1\sigma$ scatter. In the EE-11-G5 simulations, the merger time-scales converge consistently across all the mass resolutions considered.

From Fig.~\ref{fig:orb_params_fix_ecc_res_study}, we also notice that during the hardening phase, as the mass resolution increases, the spread of the eccentricities from different realizations is reduced and the eccentricity tracks also become smoother. This is consistent with the results from previous pure $N$-body simulations, i.e. as the particle mass decreases, the impact of a random star encounter on the binary becomes weaker, and thus the stochasticity of the binary eccentricity is reduced \citep[e.g.][]{Quinlan1997,Berczik2005,Nasim2020,Gualandris2022,Rawlings2023}. We further note that as the galaxy formation subgrid model contains additional stochastic processes, such as star formation (converting gas particles into stars) and SMBH accretion (swallowing gas particles), the simulations with gas can thus exhibit a higher degree of stochasticity compared to pure $N$-body simulations.

In Fig.~\ref{fig:hardening_rate_res_study}, we further investigate the dependence of binary hardening rates on particle number. The hardening rate is estimated by performing a linear fit to the inverse semimajor axis, $1/a(t)$, in the interval from $t_{\rm hard}$ to the point when $a(t) = a(t_{\rm hard})/2$. In the DD-11-G5 reruns, the hardening rates show a dependence on particle number in the low-resolution runs (i.e. with higher hardening rates for reruns with more particles), yet the Fiducial-$N_{\rm p}$ rerun converges to the Hi-res-$2N_{\rm p}$ well. Note that the particle number dependence observed in the low-resolution runs here is not related to the $N$-dependence observed in spherical galaxies from pure $N$-body simulations \citep[e.g.][]{Makino2004,Berczik2005}. In the latter case, the $N$-dependence arises from the refilling of the loss cone due to two-body relaxation, exhibiting an opposite $N$-dependence (i.e. lower hardening rates in simulations with more particles). Here, the galaxy remnant is triaxial, and the resolution dependence in the low-resolution runs originates from the galaxy formation subgrid processes. Specifically, as shown in appendix A of Paper II, the two low-resolution DD-11-G5 parent runs exhibit lower star formation rates, leading to lower central stellar densities and consequently lower hardening rates.

In the EE-11-G5 runs, there is no dependence of the hardening rates on particle number. Note that from pure $N$-body simulations, the lack of particle number influence on binary hardening rates is expected for triaxial remnants of galaxy mergers \citep{Merritt2004,Berczik2006,Khan2011,Preto2011}.

To conclude, the results (such as the binary orbital evolution and SMBH merger time-scales) from our fiducial resolution reruns, which are presented in the main text, converge well with those from the reruns conducted with a higher mass resolution.

\bsp
\label{lastpage}

\end{document}